\documentclass[useAMS,usenatbib]{mn2e}
\usepackage{amsmath,fleqn,graphicx,amssymb}
\usepackage{multirow}
\renewcommand{\[}{\begin{equation}}
\renewcommand{\]}{\end{equation}}
\def\p{\partial}\def\i{{\rm i}}
\def\Rb{R_{\rm b}}

\let\boldgrk=\gkvecten
\let\boldgrksc=\gkvecseven

\def\gkthing#1{{\mathchoice%
	{\hbox{{\boldgrk\char#1}}}
	{\hbox{{\boldgrk\char#1}}}
	{\hbox{{\boldgrksc\char#1}}}
	{\hbox{{\boldgrksc\char#1}}}}}

\def\vtheta{\gkthing{18}}

\def\vomega{\gkthing{33}}
{\newif\ifnotend
\notendtrue
\def\veclist{ABCDEFGHIJKLMNOPQRSTUVWXYZabcdefghijklmnopqrstuvwxyz.}
\def\top#1#2.{#1}
\def\tail#1#2.{#2.}
\loop\expandafter\xdef\csname v\expandafter\top\veclist\endcsname%
{{\noexpand\bf\expandafter\top\veclist}}
\edef\veclist{\expandafter\tail\veclist}
\if\veclist.\notendfalse\fi\ifnotend\repeat}
\def\d{{\rm d}}
\def\Vc{v_{\rm c}}
\def\cJ{{\cal J}}

\def\bolOm{\mbox{\boldmath$\Omega$}}
\def\vOmega{\bolOm}

\def\Myr{\,\mathrm{Myr}}
\def\kpc{\,\mathrm{kpc}}

\def\kms{\,\mathrm{km\,s}^{-1}}

\def\pc{\,\mathrm{pc}}
\def\e{\mathrm{e}}

\def\fracj#1#2{{\textstyle{#1\over#2}}}

\def\vthetaT{\vtheta^{\rm T}}\def\vJT{\vJ^{\rm T}}
\def\thetaT{\theta^{\rm T}}\def\JT{J^{\rm T}}

\def\TM{{\sc tm}}
\def\sgn{{\rm sgn}}
\def\figref#1{Fig.~\ref{#1}}

\def\vomegap{\vomega_{\rm p}}
\def\omegap{\omega_{\rm p}}

\title[Orbital tori for non-axisymmetric galaxies]
{Orbital tori for non-axisymmetric galaxies}

\author[James Binney]{
  James Binney$^1$\thanks{E-mail: binney@thphys.ox.ac.uk}\\  
  $^1$Rudolf Peierls Centre for Theoretical Physics, 1 Keble Road,
  Oxford, OX1 3NP, UK
}

\begin{document}
\maketitle

\begin{abstract}
Our Galaxy's bar makes the Galaxy's potential distinctly non-axisymmetric.
All orbits are affected by non-axisymmetry, and significant numbers are
qualitatively changed by being trapped at a resonance with the bar. Orbital
tori are used to compute these effects. Thick-disc orbits are no less likely
to be trapped by corotation or a Lindblad resonance than thin-disc orbits.
Perturbation theory is used to create non-axisymmetric orbital tori from
standard axisymmetric tori, and both trapped and untrapped orbits are
recovered to surprising accuracy. Code is added to the {\tt TorusModeller}
library that makes it as easy to manipulate non-axisymmetric tori as
axisymmetric ones. The augmented {\tt TorusModeller} is used to compute the
velocity structure of the solar neighbourhood for bars of different pattern
speeds and a simple action-based distribution function. The technique
developed here can be applied to any non-axisymmetric potential that is
stationary in a rotating from -- hence also to classical spiral structure. 
\end{abstract}


\section{Introduction} \label{sec:intro}

From numerically integrated orbits in model gravitational potentials it has
long been known that most orbits in a galaxy are quasiperiodic
\citep{BinneySI,BinneySII}. One may show that any quasiperiodic orbit is
confined to a three-dimensional torus in six-dimensional phase space
\citep{Arnold}. It is convenient to label each such torus by the Poincar\'e
invariants associated with three closed paths around the torus that cannot be
deformed into one another without leaving the torus \citep[e.g.][]{GDII}. These
labels are known as actions $J_i$, and they are singled out from other
integrals of motion, in particular energy, in that can be complemented by
canonically conjugate ``angle'' variables $\theta_i$. The latter quantify
position within a torus.

Usually orbital tori are nested one inside another, so a non-negligible
volume of phase space is filled by such a set of nested tori, and the coordinates $(\vtheta,\vJ)$
constitute an exceedingly convenient set of canonical coordinates for this
region. In certain exceptional potentials a single system of tori completely
fills phase space, so there is a global system of action-angle coordinates.
Such potentials are said to be ``integrable''. The potentials studied by
St\"ackel and named after him \citep[e.g.][]{GDII}, which yield separable
Hamilton-Jacobi equations, are the best known and most important integrable
potentials.

By Jeans' Theorem \citep{Jeans1915} the distribution function (DF) $f$ of a
steady-state galaxy can be assumed to be a function of whatever isolating
integrals are admitted by the galaxy's potential. Hence wherever in phase
space orbits are quasiperiodic, we may take the DF to be a function $f(\vJ)$
of the action integrals, and there are cogent reasons
\citep[e.g.][]{JJBPJM16} why it is advantageous to assume that $f$ depends
only on $\vJ$ rather than directly on the Hamiltonian $H(\vJ)$, as has
normally been assumed in the past.  If the galactic potential is integrable,
the DF can be a single function of $\vJ$, while if phase space breaks up into
two or more regions, each with its own sequence of nested tori, each region
will require its own functional form, and the global DF will be made up of a
patchwork of functions $f(\vJ)$, each with its own domain of validity.

In a fairly realistic Galactic potential, the orbits of some halo stars near
the Sun lie on tori that do not belong to the family that in an integrable
potential fills all phase space \citep[][hereafter B16]{JJBPJM16,Binney2016}.
B16 showed that a remarkably precise quantitative understanding of the orbits
of these stars can be achieved through the concept of resonant trapping.
Specifically, one uses the torus-mapping technique \citep{JJBPJM16} to
construct an integrable Hamiltonian that rather closely approximates the true
Hamiltonian, and then one uses first-order perturbation theory to compute the
libration of orbits around resonant orbits in the integrable
Hamiltonian. 

The trapping discussed by B16 involved the 1:1 resonance
between the radial and vertical oscillations of stars. An observational
signature of such resonant trapping has yet to be detected, but since the
work of \cite{WD98}, who used Hipparcos data to map velocity space for
stars near the Sun, it has been known that there are features in this space
that are inconsistent with the hypothesis that the DF has the form  $f(\vJ)$
with $\vJ$ the actions of an integrable axisymmetric potential.

It has long been suspected that non-axisymmetric components of the Galactic
potential are responsible for the differences between observations and the
predictions of models that assume axisymmetry
\citep{DSea04,Sellwood_LR,Hahn2011,Anea11,McMillan_Hyades,Perez2017}. Indeed,
the Galaxy is known to have a bar, that may extend as far out as $r\sim5\kpc$
\citep{BlitzSpergel1991,BGSBU,SormaniIII,Wegg2015}, and the disc is known to
carry spiral structure \citep[e.g.][]{Lepine2011}. Consequently, the
assumption of axisymmetry, which has been used with considerable success in a
large number of papers \citep{JJB12:dfs,Bovy2013,Piea14,BinneyPiffl15} cannot
be more than a starting point for a more sophisticated treatment that
includes the bar and spiral structure.

The goal of this paper is to lay the foundations for such modelling by
adapting the perturbative approach of B16 to non-axisymmetric
potentials. Specifically, a simple model of the Galactic bar is used to trap
the tori of a realistic Galactic potential around the corotation and outer Lindblad
resonances (CR, OLR). Perturbation theory yields an analytic model of the
trapped tori, so we have for these trapped tori all the functionality of tori
constructed for an axisymmetric potential by torus mapping
\citep{JJBPJM16}.

The paper is organised as follows. In Section \ref{sec:Phi} we introduce the
model of the potential of a barred Galaxy that we employ subsequently. In
Section \ref{sec:split} we explain how the Hamiltonian for motion in a
steadily rotating barred potential is decomposed into a part $H_0(\vJ)$ that
depends only on the actions and a perturbation $H_1(\vtheta,\vJ)$. In Section
\ref{sec:ptheory} we present the relevant Hamiltonian perturbation theory. In
Section~\ref{sec:Galaxy} we apply this theory to our model Galaxy, starting
in Section~\ref{sec:OLR} with the OLR, proceeding in Section~\ref{sec:CR}
to the corotation resonance, and concluding with a very brief description of
the ILR in Section~\ref{sec:ILR}. In Section~\ref{sec:Vspace} we use these results
to examine velocity space at a solar-like location within our model Galaxy.
In Section~\ref{sec:discuss} we relate our results to previous work and
consider how our results help us to understand the dynamics of a
non-axisymmetric disc. In Section~\ref{sec:conclude} we sum up and consider
directions for future work.

Appendix~\ref{sec:TMclass}  describes enhancements to the publically available code Torus
Mapper (\TM), which enable one to construct and manipulate
non-axisymmetric orbital tori, whether resonantly trapped or circulating.
TM can be can be downloaded from https://github.com/PaulMcMillan-Astro/Torus.
Appendix~\ref{sec:getV} provides the theoretical structure that \TM\ uses to
find the velocities at which a trapped torus passes through a give point.

We use throughout  Galactocentric polar coordinates $(R,z,\phi)$ with the
long axis of the bar
along $\phi=0$. Quantities referring to the azimuthal angle $\phi$ are always
listed last because in an axisymmetric potential motion in  $\phi$ is slaved
by angular-momentum conservation to the autonomous and sometimes complex
motion in the $Rz$-plane.

\section{The Galactic potential}\label{sec:Phi}

We frame our discussion in the context of a gravitational potential that
\cite{PJM11:mass} fitted to a variety of data for our Galaxy. Specifically,
we adopt the ``best'' potential in that paper, which is generated by thin and
thick stellar discs, a flattened (axisymmetric) bulge and a spherically
symmetric dark halo. Its local circular speed is $v_c=239\kms$. Our
discussion would not differ materially, however, had we adopted any
reasonably realistic axisymmetric potential. To evaluate the potential and
its derivatives we use the {\sc falPot} code distributed in the \TM\ package,
which implements an algorithm described by \cite{WDJJB98:Mass}, and was
extracted from Walter Dehnen's {\sc falcON} package
(https://github.com/Milkyway-at-home/nemo/tree/master/nemo\_cvs/usr/dehnen/falcON). 

\subsection{Non-axisymmetric component}\label{sec:nonaxi}

We add to the axisymmetric potential a non-axisymmetric perturbation, so
\[
\Phi(R,z,\phi)=\Phi_0(R,z)-\Phi_2(R,z)\cos2\phi.
\]
The non-axisymmetric contribution to the density is
\begin{align}
\rho_2(R,z,\phi)&=-(4\pi G)^{-1}\nabla^2(\Phi_2\cos2\phi)\nonumber\\
&=-(4\pi G)^{-1}\left(\nabla^2\Phi_2-4\Phi_2/R^2\right)\cos2\phi.
\end{align}
This should vanish at both small and large radii. In fact, as the $z$ axis is
approached  it must vanish at least as fast as $R^2$. At large $R$ it should
tend to the external potential of a quadrupole, so decay like $r^{-3}$. These
conditions are satisfied by the ansatz
\[\label{eq:Phi2}
\Phi_2(R,z)={KR^2\over (\Rb^2+m^2)^{5/2}},
\]
 where $\Rb$ is a scale radius,
\[
m^2\equiv R^2+{z^2\over q^2},
\]
  and the constant $q$ controls the flatness of the non-axisymmetric
contribution to the density. The normalising constant $K$ in equation
(\ref{eq:Phi2}) has dimensions of velocity squared times
length cubed, so it is convenient to control its value through the
dimensionless number
\[
A\equiv {K\over v_{\rm c}^2R_{\rm b}^3},
\]
 where $v_{\rm c}$ is the circular speed at the solar radius that is implied by the
axisymmetric component of the potential.

\begin{figure}
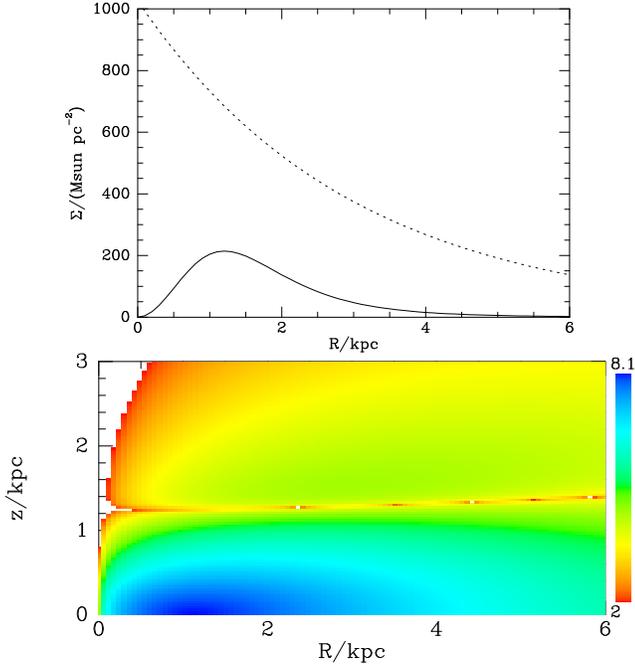

\begin{center}
\includegraphics[width=.8\hsize]{figs/Sigs.ps}
\includegraphics[width=.99\hsize]{figs/xz.ps}
\end{center}
\caption{Top: the full curve shows the surface density $\Sigma(R)=\int\d z\,\rho_2(R,z)$ that
generates  the non-axisymmetric component of the potential (\ref{eq:Phi2})
with $q=0.9$, while the dotted curve shows the axisymmetric density to which
this is added. Bottom: the density $\rho_2(R,z)$ for the same value of
$q$. The colours indicate values of $\log_{10}\rho_2$.  Above the line of
three red dots the density is
negative.}\label{fig:Phi2plot}
\end{figure}

\begin{figure*}
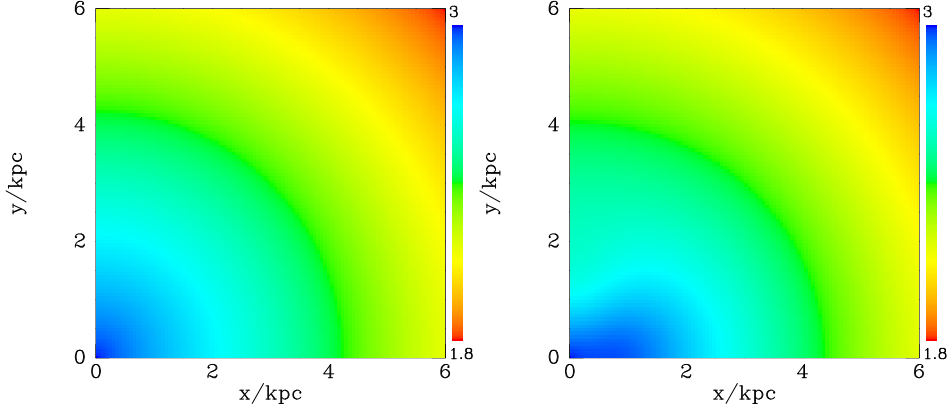

\begin{center}
\includegraphics[width=.35\hsize]{figs/xy0.ps}\ 
\includegraphics[width=.35\hsize]{figs/xy.ps}
\end{center}
\caption{Right: the surface density obtained on adding the non-axisymmetric
component plotted in \figref{fig:Phi2plot} to the axisymmetric surface
density plotted in the left panel. The latter has exponential scale length
$R_\d=3\kpc$.}\label{fig:addSD}
\end{figure*}

We now have
\[
{\d\Phi_2\over\d R}=KR{2\Rb^2-3R^2+2z^2/q^2\over(\Rb^2+m^2)^{7/2}},
\]
so 
\begin{align}
{1\over R}{\d\over\d R}&\left(R{\d\Phi_2\over\d R}\right)
=K\biggl[2(2\Rb^2-6R^2+2z^2/q^2)(\Rb^2+m^2)\nonumber\\
&-7R^2(2\Rb^2-3R^2+2z^2/q^2)\biggr]\Big/(\Rb^2+m^2)^{9/2}.
\end{align}
Similarly
\[
{\d\Phi_2\over\d z}=
-5K{R^2\over q^2}{z\over(\Rb^2+m^2)^{7/2}},
\]
and 
\[
{\d^2\Phi_2\over\d z^2}=
5K{R^2\over q^2}
\,{7z^2/q^2-(\Rb^2+m^2)\over(\Rb^2+m^2)^{9/2}}.
\]
From these formulae the density $\rho_2(R,z)$ is readily computed. 

\subsection{Normalisation of the bar}

\cite{SormaniIII} simulated the flow in two dimensions of isothermal gas
flow in model Galactic potentials and by comparing the simulations to
radio-frequency observations of Galactic gas estimated the pattern speed and
amplitude of the bar. They concluded that the bar has to generate a
substantial quadrupole moment. We fix the strength of our bar by reference to
that of their most successful models.

Whereas we have chosen to work from a simple analytic function for the
coefficient of proportionality, $\Phi_2(R,z)$, \cite{SormaniIII} assumed a
particular form for the corresponding density:
\[\label{eq:Srho}
4\pi G\rho_2(\vx)=A_{\rm S}\left({v_0\e\over
r_q}\right)^2\e^{-2r/r_q}\sin^2\theta\cos2\phi,
\] 
 where $v_0=220\kms$, $r_q=1.5\kpc$ and $A_{\rm S}\ga0.4$ is a dimensionless
amplitude.

The strength of a bar is quantified by  $\lim_{r\to\infty}r^3\Phi_2(\vr)$.
Since $\sin^2\theta\cos2\phi=\sqrt{8\pi/15}\,(Y_2^2+Y_2^{-2})$, in the
notation of equation (2.95) in \cite{GDII} at large $\vr$
\[
4\pi G\rho_{22}=A_{\rm S}\sqrt{8\pi\over15}\left({v_0\e\over
r_q}\right)^2\e^{-2r/r_q}.
\]
 Consequently, at large $r$ the potential generated by (\ref{eq:Srho}) is
\[
\Phi_2(\vx)=A_{\rm S}\sqrt{8\pi\over15}{Y_2^2+Y_2^{-2}\over5r^3}
\left({v_0\e\over
r_q}\right)^2\int_0^r\d a\,a^4\e^{-2a/r_q}
\]
For $r\gg r_q$ the integral tends to $4!(r_q/2)^5$, so
\begin{eqnarray}
r^3\Phi_2(\vx)&=&{4!\e^2A_{\rm S}\over5\times32}\sqrt{8\pi\over15}\,(Y_2^2+Y_2^{-2})
v_0^2r_q^3\nonumber\\
&=&\frac{3\e^2A_{\rm S}}{20}v_0^2r_q^3\sin^2\theta\cos2\phi.
\end{eqnarray}

When $z=0$ so $m=R=r$, our expression (\ref{eq:Phi2}) for $\Phi_2$ yields
$r^3\Phi_2(r,0)\to K=Av_{\rm c}^2R_{\rm b}^3$, so to achieve the same quadrupole as \cite{SormaniIII}
we must set
\[
A=\frac{3\e^2}{20}\left({v_0\over v_{\rm c}}\right)^2
\left({r_q\over R_{\rm b}}\right)^3A_{\rm S}.
\]
In this formula, we adopt $\Vc=239\kms$ and $A_{\rm S}=0.4$.

It remains to choose the bar's scale length, $\Rb$.  Guided by
plots like those of Figs.~\ref{fig:Phi2plot} and \ref{fig:addSD}, we adopt
$\Rb=2.09\kpc$, which is $0.7$ times the density weighted disc scale lengths
of the axisymmetric component. Then the figures show that the surface density
is strongly barred inside $R=3\kpc$ and significantly non-axisymmetric out to
$R=4\kpc$.  Vertically the bar extends to $|z|\sim1\kpc$. 

\subsection{Pattern speed}

\cite{SormaniIII} concluded that the bar's pattern speed is
$\omegap\simeq40\kms\kpc^{-1}$, a value that coincides nicely the estimate
obtained by \cite{Wegg2015} by modelling the kinematics of bulge stars.
\cite{Perez2017} show that $\omegap=39\kms$ correctly predicts the kinematics
of solar-neighbourhood stars because it causes stars trapped by corotation to
visit the solar neighbourhood in the form of the ``Hercules stream''. Over
the previous two decades larger values $\sim55\kms\kpc^{-1}$ were favoured,
in part on account of a significant under-estimate of the length of the bar
by \cite{BGSBU} and in part through the Hercules stream being supposed to
arise by trapping at the OLR rather than corotation
\citep{WD98,MonariStayAway2017}. In the following we adopt
$\omegap=0.04\Myr^{-1}=39\kms\kpc^{-1}$.

We take the pattern speed to be positive, so the bar rotates anticlockwise in
the $xy$ plane. We usually consider that our Galaxy rotates clockwise because
we imagine viewing it from the Northern hemisphere. To apply the present
model to our Galaxy we must view the latter from Australia. The end of the
bar that is nearer to the Sun is rotating away from the Sun. We achieve a
similar location for a model Sun by placing it at $\phi=155$ degrees.

\section{Splitting the Hamiltonian}\label{sec:split}

Our goal is to handle perturbations to an axisymmetric system that rotates at
a constant angular velocity $\vomegap$. Let $\vv$ be the velocity of a star
in the frame  of reference that rotates with the system, then the star's kinetic energy is
\[
K=\fracj12|\vv+\vomegap\times\vx|^2,
\]
so the Lagrangian is
\[\label{eq:Lag}
L=\fracj12|\vv+\vomegap\times\vx|^2-\Phi(\vx),
\]
 where the original time dependence of the gravitational potential $\Phi$ has
 been absorbed into the rotation of the reference frame. 
It follows from equation (\ref{eq:Lag}) that the canonical  momentum is
\[
\vp={\p L\over\p\vv}=\vv+\vomegap\times\vx,
\]
which is in fact the momentum in the underlying inertial frame. The
Hamiltonian is
\[\label{eq:H}
H(\vx,\vp)=\vp\cdot\vv-L=\fracj12|\vp|^2+\Phi-\vomegap\cdot(\vx\times\vp).
\]
Since $\vp$ is the inertial-frame momentum, the first two terms on the right
of equation (\ref{eq:H}) comprise the star's energy in
the inertial frame, and the final term is simply $\omegap p_\phi$, where
$p_\phi$ is the component of angular momentum parallel to $\vomegap$. Hence
we can write
\[\label{eq:Hrot}
H(\vx,\vp)=H_{\rm in}(\vx,\vp)-\omegap p_\phi,
\]
 where $H_{\rm in}$ is the Hamiltonian that would apply if the  given
potential were not rotating but stationary in an inertial frame. The shift
from $H_{\rm in}$ to $H$ that encodes  rotation of the potential changes
only one equation of motion, namely that of $\phi$:
\[
\dot\phi=[\phi,H]={\p H_{\rm in}\over\p p_\phi}-\omegap.
\]
  In the rotating frame, $\phi$ increments more slowly than in the
inertial frame because the zero-point of angle is advancing at the rate
$\omegap$.

By torus mapping  it will usually be possible to construct a Hamiltonian
$\overline{H}(\vJ)$ that admits a global system of angle action coordinates
$(\vtheta,\vJ)$ and is very close to $H_{\rm in}$. That is, we can write
\[\label{eq:Hin}
H_{\rm in}(\vtheta,\vJ)=\overline{H}(\vJ)+H_1(\vtheta,\vJ)
\]
 where $H_1\ll \overline{H}$. If $\Phi$ is not very strongly non-axisymmetric, we
can map tori in such a way that $\overline{H}$ is axisymmetric, so $p_\phi=J_\phi$ is
one of the actions in the set $\vJ$. In this case, the true Hamiltonian 
\[\label{eq:Hrot2}
H(\vtheta,\vJ)=\bigl\{\overline{H}(\vJ)-\omegap J_\phi\bigr\}+H_1(\vtheta,\vJ),
\]
 has been successfully split into a dominant  part
\[
H_0(\vJ)\equiv \overline{H}(\vJ)-\omegap J_\phi
\]
 that does not depend on
the angle variables and a small perturbation $H_1$.

It will sometimes be expedient to map tori in such a way that
$\overline{H}(\vJ)$ is non-axisymmetric. If we do map tori thus, we should
be able to arrange for $\overline{H}(\vJ)$ to provide a closer approximation
to $H_{\rm in}$, and thus make $H_1(\vtheta,\vJ)$ smaller than if we keep
$\overline{H}(\vJ)$ axisymmetric. The disadvantage
of making $\overline{H}$ non-axisymmetric is that then $p_\phi$ will not
coincide with any of the actions in the set $\vJ$ -- the way each action
depends on $(\vx,\vv)$ depends on the potential $\Phi(\vx)$, and when $\Phi$
is non-axisymmetric $p_\phi=R^2\dot\phi$ will not be an action variable.
However, as the degree of non-axisymmetry vanishes, one of the actions, which
we may call $J_\phi$, will converge on $p_\phi$. With a non-axisymmetric
$\overline{H}(\vJ)$, the $\omegap p_\phi$ term
in the full Hamiltonian will contribute to the perturbative part of the
Hamiltonian alongside $H_1$. That is, in this case we must write
\[
H(\vtheta,\vJ)=\left\{\overline{H}(\vJ)-\omegap J_\phi\right\}
+\left[H_1(\vtheta,\vJ)+(J_\phi-p_\phi)\omegap\right],
\]
 such that the entire square bracket is a perturbation on the integrable
Hamiltonian that is defined by the curly bracket. The residual $H_1$ from
\TM\ will be substantially smaller than the full non-axisymmetric component
of the potential with which we will contend, but this gain will be to some
extent offset by the appearance of the perturbation $(J_\phi-p_\phi)\omegap$.
We leave exploration of this approach to a later study.

\section{Resonance}\label{sec:ptheory}

Now we present the theory of resonant trapping. This theory relates to the
state of {\it being} trapped by a resonance and does not address the complex process
by which a star {\it becomes} trapped, for example as a bar strengthens and
slows down, with the consequence that it sweeps into its embrace stars that
previously orbited freely outside it.
There is significant overlap
between this section and Section 5 of B16, but there are some
subtle differences and we need to establish notation for subsequent use.

\subsection{Pendulum dynamics}\label{sec:pendulum}

A resonant orbit in $H_0(\vJ)$ is one on which
\[\label{eq:rescon}
\vN\cdot\vOmega=0,
\]
where $\vN$ is a
vector with integer components and
\[
\vOmega\equiv{\p H_0\over\p\vJ}
\]
 is the vector of frequencies.  In the rotating frame,
$\Omega_\phi$ changes sign at corotation from positive inside corotation to
negative outside it. Hence to satisfy equation (\ref{eq:rescon}) with
$N_z=0$, the ratio $N_r/N_\phi$ has to rise from  negative values  inside
corotation, through zero at corotation to positive values outside corotation.
The best known resonances of this class are the inner Lindblad resonance,
$\vN=(1,0,-2)$, the ultraharmonic resonance $\vN=(1,0,-4)$, the
corotation resonance $\vN=(0,0,1)$ and the outer
Lindblad resonance, $\vN=(1,0,2)$.
B16 examined the case $\vN=(1,1,0)$ and we can exploit the formalism
developed there by simply replacing $\overline{H}$ in the formulae of B16
by $H_0$.

Near a resonance, the angle
variable
\[\label{eq:thetas}
\theta_1'\equiv\vN\cdot\vtheta
\]
 evolves slowly and we make a canonical transformation to new
angle-action variables $(\vtheta',\vJ')$ that include this variable. We use
the generating function
\[
S'(\vtheta,\vJ')=J_1'\vN\cdot\vtheta+J_2'\theta_2+J_3'\theta_3,
\]
 which ensures  that
equation \eqref{eq:thetas} holds and makes
$\theta'_{2,3}=\theta_{2,3}$. From $\vJ=\p S/\p\vtheta$ we find
\begin{align}\label{eq:defsJp}
J_1&=N_1J'_1     &                 J'_1&=J_1/N_1\cr
J_2&=N_2J'_1+J'_2&\leftrightarrow\qquad  J'_2&=J_2-J_1{N_2/N_1}\cr
J_3&=N_3J'_1+J'_3&                 J'_3&=J_3-J_1{N_3/N_1}.
\end{align}

We Fourier expand the Hamiltonian in the new angle variables
\[\label{eq:expH}
H(\vtheta',\vJ')=H_0(\vJ')
+\sum_{\vk\ne0}\hat h_\vk\e^{\i\vk\cdot\vtheta'},
\]
 where $|\hat h_\vk|\ll H_0$.
The new actions have the equations of motion
\[\label{eq:fullJdot}
\dot\vJ'=-{\p H\over\p\vtheta'}=-\i\sum_\vk \vk
\hat h_\vk\e^{\i\vk\cdot\vtheta'}.
\]
 Averaging these equations over the fast angles $\theta_{2,3}'$ we conclude
that the actions $J_{2,3}'$ are effectively constant under the perturbation,
so we need consider only the system's motion in the $(\theta_1',J_1')$ plane.
This motion is governed by the Hamiltonian
\[\label{eq:Hbar}
H(\theta'_1,\vJ')=H_0(J_1')
+\sum_{n\ne0}\hat h_n(J'_1)\e^{\i n\theta'_1},
\]
 where $\hat h_n\equiv\hat h_{(n,0,0)}$ and we have omitted references to the constant
actions $J'_{2,3}$. Since this a time-independent Hamiltonian, the motion
occurs on the curve in the $(\theta'_1,J'_1)$ plane on which $H=I$, a constant. A
good approximation to this motion can be obtained by Taylor expanding the
functions of $J'_1$ in equation \eqref{eq:Hbar}. However, before we do so we
exploit the
reality of $H$ to write
\[\label{eq:H1d}
H(\theta'_1,\vJ')\simeq H_0(J_1')
+2\sum_nh_n(J'_1)\cos(n\theta'_1+\psi_n),
\]
 where the $h_n$ are the amplitudes and $\psi_n$ are the phases of the
 complex variables $\hat h_n$. We expand $H_0$ and $h_n$ to second
order in 
\[
\Delta\equiv J'_1-J'_{01},
\]
 where $J'_{01}$ is the primed action
of the resonant torus. Since a constant term in $H$ can be discarded and we
know that $\p H_0/\p J'_1=0$ on the resonant torus
(eq.~\ref{eq:rescon}), we replace
$H_0$ by $\fracj12 G\Delta^2$, where
\[
G\equiv{\p^2H_0\over\p J'_1{}^2}={\p{\Omega}'_1\over\p
J'_1}.
\]
The Taylor series for $h_n(J'_1)$,
\[
h_n(J'_1)=h_n^{(0)}+h_n^{(1)}\Delta+\fracj12 h_n^{(2)}\Delta^2+\cdots
\]
cannot be simplified in this way, so the
equation $H=I$ becomes
\begin{align}\label{eq:Hquad}
0&=\Bigl(\fracj12G+\sum_nh_n^{(2)}\cos(n\theta'_1+\psi_n)\Bigr)\Delta^2\cr
&+2\Bigl(\sum_nh_n^{(1)}\cos(n\theta'_1+\psi_n)\Bigr)\Delta\cr
&+\Bigl(2\sum_nh_n^{(0)}\cos(n\theta'_1+\psi_n)-I\Bigr).
\end{align}
In this approximation we can determine $J'_1(\theta'_1)$ simply by solving
the quadratic equation \eqref{eq:Hquad} for $\Delta$ given $\theta'_1$. Given the
oscillating value of $J'_1$ and the constants $J'_{2,3}$, we can recover
the complete action-space coordinates from  equations \eqref{eq:defsJp}. In
general all three components of $\vJ$ oscillate but in such a way that $H$ is
to leading order constant.

If we retain only one value of $n$ and neglect $h_n^{(1)}$ and $h_n^{(2)}$,
equation \eqref{eq:Hquad} reduces to the energy equation of a pendulum, and
this is traditionally used to discuss resonant trapping
\citep[e.g.][]{Chirikov1979}. \cite{Kaasalainen_res} demonstrated the merit
of retaining $h_n^{(1)}$ and $h_n^{(2)}$. B16 retained two values of $n$, but
we shall find in the applications considered here that there is only one
value of $n$ to consider.

\subsection{Action and angle of libration}

The range through which $\theta'_1$ oscillates is set by the condition that
the quadratic for $\Delta$ has real roots:
\begin{align}\label{eq:discr}
\Bigl[\sum_nh_n^{(1)}\cos(n\theta'_1+\psi_n)\Bigr]^2&\ge
\Bigl(\fracj12G+\sum_nh_n^{(2)}\cos(n\theta'_1+\psi_n)\Bigr)\cr
&\times\Bigl(2\sum_nh_n^{(0)}\cos(n\theta'_1+\psi_n)-I\Bigr).
\end{align}
 When $G<0$, as is the case at both OLR and CR, condition (\ref{eq:discr}) is
satisfied for the widest range of angles when $I<0$ also, so the second
bracket on the right is often positive.  If $I$ is less than a critical value
$I_{\rm bot}$, the condition is satisfied but for no value of $\theta_1'$
becomes by equality. Consequently, at no value of $\theta_1'$ do the roots
coincide, and the orbit is restricted to one root or the other.  This is the
regime of circulation. The two roots corresponds to circulation inside and
outside the region of entrapment. 

When $I$ is larger than $I_{\rm bot}$, condition (\ref{eq:discr}) is
satisfied in a restricted range in $\theta_1'$, and an orbit can transfer
from one root to the other at the extremes of this range. This
is the regime of libration. As $I$ increases, the range shrinks and it vanishes
at $I=I_{\rm top}$. When $I=I_{\rm top}$ the amplitude of libration vanishes.

In the case of the OLR we find that  $G<0$ and $h_n$ vanishes for $n>1$ and that
$\psi_1=\pi$.  So when $h^{(1)}$ and $h^{(2)}$ are
neglected, the libration condition is
\[\label{eq:libcond}
I\le2h_1^{(0)}\cos(\theta_1'+\pi).
\]
Given that the $h_n$ are by definition all non-negative, it follows that
libration occurs around $\theta_1'=\pi$.  Hence, when $I=I_{\rm
top}\simeq2h_1^{(0)}$, equation (\ref{eq:libcond}) forces $\theta_1'=\pi$, and
the libration amplitude vanishes. Conversely, when $I=I_{\rm
bot}\simeq-2h_1^{(0)}$ the libration amplitude becomes $\pi$.  It follows
that $I_{\rm bot}$ is the value of $I$ for which condition (\ref{eq:discr})
is an equality with the cosines set to $-1$, while $I_{\rm top}$ is the value
of $I$ at which equality is reached with the cosines equal to $+1$.

Each value of $I$ in the range $I_{\rm bot}<I\le I_{\rm top}$ corresponds to an action $\cJ$ that quantifies the
extent to which a trapped orbit oscillates around the trapping torus. $\cJ$
is straightforwardly computed as
\[\label{eq:resJ}
\cJ={1\over2\pi}\oint\d\theta'_1 J'_1(\theta'_1),
\]
 where the dependence of $J'_1$ on $\theta'_1$ is obtained from equation
\eqref{eq:Hquad}. Since $\theta'_1$ increases from its minimum to its maximum
value with $\Delta$ given by the larger root of the quadratic
\eqref{eq:Hquad} and returns to its minimum value with $\Delta$ given by the
smaller root, $\cJ$ is give by the difference of the roots $\Delta$
integrated over the range of $\theta'_1$.

The procedure for computing the actions of circulating orbits is very
similar: one averages the relevant root $\Delta(\theta_1')$ with respect to
$\theta_1'$ and then adds
this average to the value of $J_1'$ for the perfectly resonant torus to
obtain the value of $J_1'$ on the newly created torus.

On a trapped or nearly trapped orbit, $\theta'_1$ is not an angle variable, although
$\theta'_{2,3}$ are angle variables. Since the missing angle variable evolves
linearly in time, it is
\[
\vartheta(\theta'_1)=2\pi{\int_0^{\theta'_1}\d\theta'_1/\dot\theta'_1\over
\oint\d\theta'_1/\dot\theta'_1},
\]
 where from Hamilton's equation and equation  \eqref{eq:H1d} we have
\[
\dot\theta'_1=G\Delta+2
\sum_n\left(h_n^{(1)}+h_n^{(2)}\Delta\right)\cos(n\theta'_1+\psi_n).
\]
 In this equation $\Delta$ is by the quadratic equation \eqref{eq:Hquad} a function
of $\theta'_1$.

\begin{figure}
\includegraphics[width=\hsize]{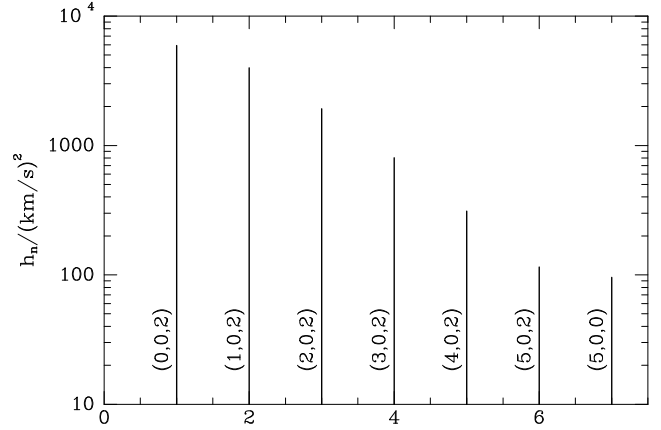}
\caption{The amplitudes of the Fourier components of $H$ on a torus that
satisfies the condition for the OLR. The six largest terms all belong to a
single sequence $\vk=(j,0,2)$ for $j=0,1,\ldots$.}\label{fig:nonres}
\end{figure}

\subsection{Impact of non-resonant terms}

Above we neglected  all non-resonant terms by averaging the
equations of motion (\ref{eq:fullJdot}) over the fast angles. This procedure
works extremely well when the tori are fitted to the full Hamiltonian rather
than just a symmetric part of it because the fitting procedure can, in
principle, reduce the non-resonant terms in the Fourier decomposition of $H$
to arbitrarily small values. Unfortunately, in the present case the torus
is fitted to the axisymmetric part of $H$ and the variation of H over the
fitted tori include significant non-resonant terms. We should
include these terms in the perturbative treatment.

Fig.~\ref{fig:nonres} illustrates this point by showing the values of the
$h_\vk$ for a torus on which the condition for the OLR is perfectly
satisfied. The largest term is non-resonant, being proportional to
$\cos2\theta_\phi$. There follow a series of terms that are proportional to
$\cos(n_r\theta_r+2\theta_\phi)$. Of these terms only the largest, which has
$n_r=1$, is resonant. The impact of this term was computed above. Here we
compute the impact of the non-resonant terms.

It is convenient to work in the $(\vtheta',\vJ')$ coordinate system in which,
according to the theory of Section \ref{sec:pendulum}, $J_2'$ and $J_3'$ are constant,
while $J_1'$ obeys a pendulum equation. The non-resonant terms induce small
oscillations in $J_3'$ and slightly modify the pendulum motion of $J_1'$.

We consider first the largest non-resonant term, which is a particularly
simple case because it has $n_r=0$ and 
the only equation of motion affected by this term is that for $J_3'$.
Moreover $\theta_3'=\theta_\phi$, so we
may write
\[
{\d J_3'\over\d\theta_3'}={\dot J_3'\over\Omega_3'}
=-{1\over\Omega_3'}{\p H\over\p\theta_3'}.
\]
Integrating this equation, it follows that 
\[\label{eq:non-res}
J_3'(\theta_3')=\overline{J_3'}-{2h_{002}\over\Omega_3'}\cos(2\theta_3'+\psi_{002}),
\]
 where $\overline{J_3'}$ denotes the average value of $J_3'$, which we
interpret to be value of $J_3'$ for the perfectly resonant torus. It is
perhaps worth noting that equation (\ref{eq:non-res}) could have been
obtained by requiring that at any point on the orbit the change in the
unperturbed Hamiltonian,
\[
\delta H_0={\p H_0\over\p J_3'}\,\delta J_3',
\]
cancel the numerical value of the perturbing Hamiltonian.

In a similar vein, the contributions of the other non-resonant terms to the
equations of motion are
\begin{eqnarray}\label{eq:non-res2}
\delta J_1'&=&-\int{\d\theta_3'\over\Omega_3'}{\p\delta
H\over\p\theta_1'}\nonumber\\
&=&-\int{\d\theta_3'\over\Omega_3'}{n/N_1\over2-nN_3/N_1}{\p\delta
H\over\p\theta_3'}\nonumber\\
&=&-{2nh_{n02}\over\Omega_3'(2N_1-nN_3)}\cos(n\theta_r+2\theta_\phi+\psi_{n02})\\
\delta J_3'&=&-\int{\d\theta_3'\over\Omega_3'}{\p\delta
H\over\p\theta_3'}\nonumber\\
&=&-{2h_{n02}\over\Omega_3'}\cos(n\theta_r+2\theta_\phi+\psi_{n02})\nonumber
\end{eqnarray}
 When we set $n=0$ and $N_3=2$ these equations are equivalent to equation
(\ref{eq:non-res}) for the change induced by the largest perturbation.

For each relevant value of $n=2,3,\ldots$ we increment $J_1'$ and $J_3'$ by
the amounts given in equation (\ref{eq:non-res2}) from the values yielded by
the resonant theory of Section \ref{sec:pendulum}. 

\subsection{Resonant or non-resonant theory?}\label{sec:resornonres}

The point of the pendulum equation is that it yields an accurate
representation of the non-uniform evolution of the resonant angle
$\theta_1'=\vN\cdot\vtheta$. The equation's weakness is that it derives its
data from the immediate vicinity of the perfectly resonant torus.
Specifically, we use with it values of $G$ and the resonant Fourier
amplitudes $h_n$ that were obtained from Taylor series, but the employed
values of the non-resonant amplitudes $h_\vk$ are simply those of the
perfectly resonant torus. Near that torus, the pendulum equation provides
accurate results, but as we move away, errors arising from stale values of
$h_\vk$ grow. Meanwhile, as we move deeper into the region of circulation,
the motion of the resonant angle becomes faster and more uniform. At some
point it becomes advantageous to abandon the pendulum equation and simply
apply equations (\ref{eq:non-res2}), but now including {\it all} the $h_\vk$
rather than just the non-resonant terms, and using the values of $h_\vk$ and
$\vOmega$ for the current torus rather than for the perfectly resonant torus.

\begin{figure}
\includegraphics[width=\hsize]{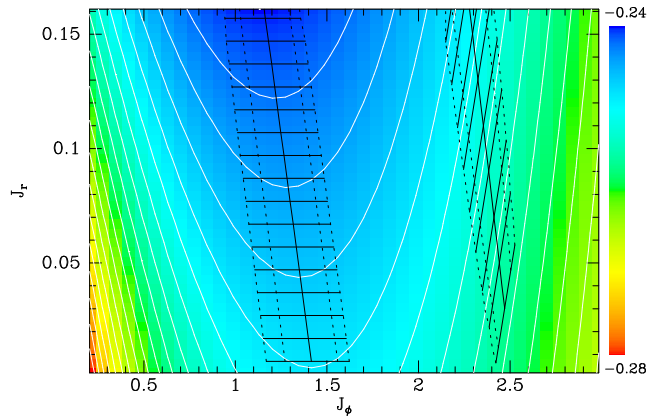}
 \caption{The colour scale and white contours give the value of the
axisymmetric Hamiltonian $H_0$ in the $(J_\phi,J_r)$ plane at
$J_z=0.0025\kpc^2\Myr^{-1}$. The zones of entrapment at corotation (centre)
and at the OLR (right) are marked by black ladder-like structures. A full line down
the middle of each structure joins unperturbed tori on which the resonance
condition is satisfied. The dashed lines on each side mark the boundaries of
the region within which orbits are trapped computed from perturbation theory.
The actions of trapped orbits oscillate along lines parallel to the rungs of
each ladder. Along the lines on which the resonance condition is satisfied, the
rungs are tangent to contours of constant $H_0$. The shell orbit
through the Sun sits near
the $x$ axis at $J_\phi=1.96\kpc^2\Myr$.}\label{fig:Hbar}
\end{figure}

\section{Examples}\label{sec:Galaxy}

In this section we show results obtained when the formalism of the last
section is applied to the Hamiltonian associated with the model Galactic
potential of Section~\ref{sec:split}.  In Fig.~\ref{fig:Hbar} the colour
scale and white contours show the value of the axisymmetric Hamiltonian
$H_0(\vJ)$ through a significant part of action space. At a given
(small) value of $J_r$, $H_0$ peaks at the value of $J_\phi$
associated with corotation. This ridge-line in $H_0$ corresponds to
the fact that in a rotating frame the effective potential
$\Phi(R,z)-\fracj12\omegap^2R^2$ has the structure of a volcano in that it
falls away from a roughly circular rim in the vicinity of corotation into the
crater on one side and down towards the flatlands on the other
\citep[e.g.][Figure 3.14]{GDII}. Fig.~\ref{fig:Hbar} is computed for vertical
action $J_z=0.0025\kpc^2\Myr^{-1}$, which typically lets solar-neighbourhood
orbits reach $z\sim0.27\kpc$. However, the diagram computed for
$J_z=0.025\kpc\Myr^{-1}$, which lets stars reach $z\sim0.97\kpc$, is virtually
indistinguishable. Thus the response of stars to the bar will not depend on
whether they belong to the thin or thick disc.

Two black ladder-like structures are evident in Fig.~\ref{fig:Hbar}. The left-hand
ladder marks the region within which stars are trapped at corotation, while
the right-hand ladder marks the region of entrapment by the OLR. The line
down the centre of each ladder marks the tori on which the relevant resonance
condition is exactly satisfied. The shorter lines or ``rungs'' that cross
this central line are the tangents to the contours of constant $H_0$
where they cross the central line. During libration the locations of orbits
in the $(J_\phi,J_r)$ plane oscillate parallel to these rungs. Each side of
the ladder is marked by two parallel lines. The distance across a ladder between the
inner lines measures the maximum amplitude of libration of trapped orbits,
which is simply the value of the libration action $\cJ$ obtained by setting
$I=I_{\rm bot}$ in the pendulum formulae.  The outer lines indicate the
furthest excursions $\Delta$ of trapped orbits from the actions of the underlying
perfectly resonant orbit.  Within the phase space associated with the gap
between the inner and outer lines there are orbits that librate and orbits
that circulate. These orbits are all best computed with resonant perturbation
theory, while outside this region non-resonant theory is usually more
reliable.

\begin{figure}
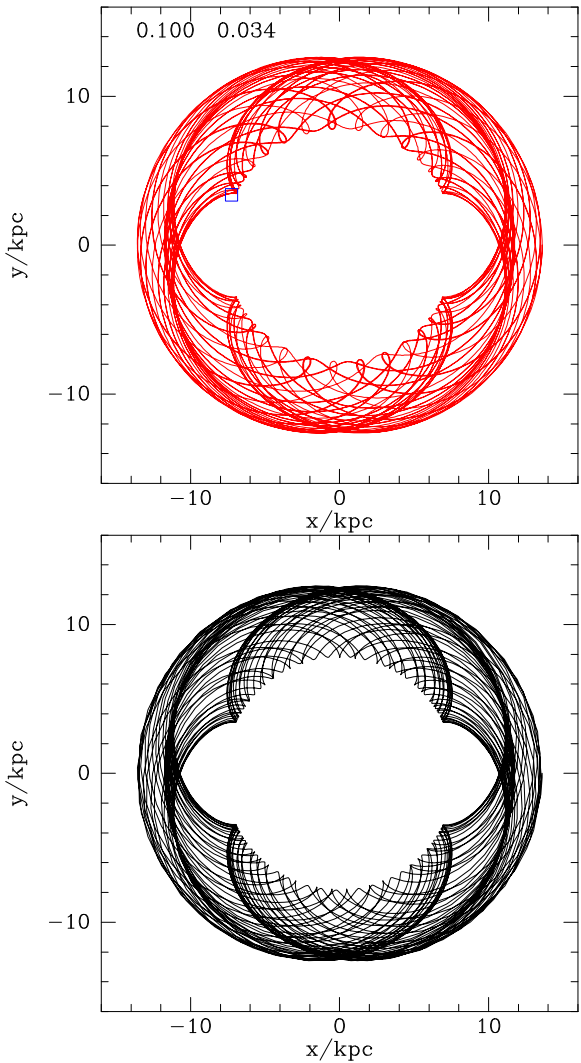

\begin{center}
\centerline{\includegraphics[width=.9\hsize]{figs/OLRred.ps}}
\centerline{\includegraphics[width=.9\hsize]{figs/OLRblack.ps}}
\end{center}
\caption{In red an orbit trapped at OLR generated with perturbation theory.
In black the same orbit integrated numerically. For this orbit
$J_z=0.0025\kpc^2\Myr^{-1}$,
$J_3'=2.17\kpc^2\Myr^{-1}$,
$I=0.6I_{\rm bot}$ and $\cJ=0.034\kpc^2\Myr$. The orbit reaches distance
$|z|=0.43\kpc$ from the plane. The underlying resonant orbit has
$J_r=0.1\kpc^2\Myr^{-1}$. The black square in the upper panel shows the likely location of the
Sun.}\label{fig:overplot}
\end{figure}

\begin{figure}
\includegraphics[width=\hsize]{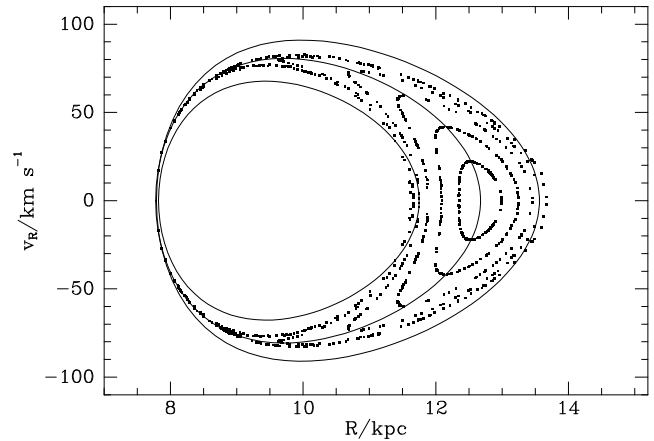}
\caption{A surface of section $\phi=z=0$. Each  curve is a cross-section
through a torus of the axisymmetric Hamiltonian $H_0$. The resonance
condition for the OLR is satisfied on the middle torus. The points are the
consequents of orbits integrated in the full barred Hamiltonian from points
on the resonant torus.}\label{fig:no_fit}
\end{figure}

\begin{figure}
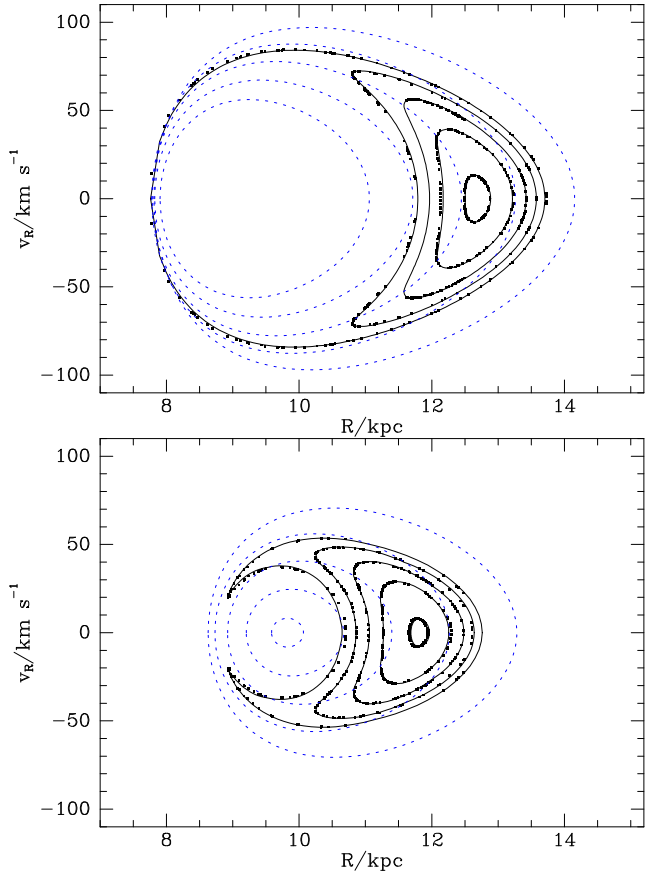

\includegraphics[width=\hsize]{figs/sosOLR1.ps}
\includegraphics[width=\hsize]{figs/sosOLR2.ps}
 \caption{Two surfaces of section $\phi=z=0$. The full black curves are cross
sections through tori constructed perturbatively. The points are consequents
on orbits started from a point on each of these curves. The broken blue
curves show cross sections through unperturbed tori like those used to
construct the perturbed tori. The upper panel is for the case that the
underlying resonant orbit has $J_r=0.1\kpc^2\Myr^{-1}$ while the lower panel
is for $J_r=0.035\kpc^2\Myr^{-1}$. In the latter case the trapping region
approaches $J_r=0$, so the innermost broken blue curve is very
small. In the upper panel the lowest value of $I=1.05I_{\rm bot}$ so the
orbit generating the outermost curve circulates, while in the lower panel the
smallest value of $I=0.98I_{\rm bot}$ so the outermost orbit
librates.}\label{fig:sos2}
\end{figure}

\subsection{Trapping at the OLR}\label{sec:OLR}

Fig.~\ref{fig:overplot} shows a typical orbit trapped at the OLR. The red
curve shows the trajectory yielded by perturbation theory, while the
black curve shows the trajectory obtained by numerically
integrating the full equations of motion from a single phase-space point on
the red trajectory. Since perturbation theory yields a close approximation to
an orbit, starting the black trajectory from a different point makes very
little difference to the figure.

Fig.~\ref{fig:no_fit} is  a surface of section $\phi=z=0$ at
$J_z=0.0025\kpc^2\Myr^{-1}$ and $J_3'=2.17\kpc^2\Myr^{-1}$. The full curves
are cross sections through tori in the axisymmetric Hamiltonian
$H_0$ obtained by torus mapping \citep{JJBPJM16}. The condition
for resonance at the OLR is satisfied on the middle torus, and the tori
generating the outermost and innermost curves are the tori that perturbation
theory predicts will bound the
region of entrapment. 

The points in Fig.~\ref{fig:no_fit} are consequents of seven orbits
integrated with the full equations of motion from initial conditions on the
resonant torus. The consequents of any given orbit lie on a curve because
these are quasiperiodic orbits. But for only two orbits does this curve
resemble the cross section of a torus generated by \TM. It is evident that on
the orbits that gave rise to the innermost and outermost sets of consequents
the resonant angle $\theta_1'$ circulates, while in the intervening five
orbits $\theta_1'$ librates. We see that the boundaries of the region of
entrapment are quite accurately predicted by perturbation theory.  The cross
sections of the tori almost coincide at small $R$ but this is an illusion:
the tori have distinct values of $J_\phi$, with the consequence that at a
given $R$ they are well separated in $v_\phi$. 

The upper panel of Fig.~\ref{fig:sos2} shows the same surface of section as
Fig.~\ref{fig:no_fit} but with the cross sections through tori of
$H_0$ now shown in broken blue lines.  The full black curves in
Fig.~\ref{fig:sos2} are cross sections through tori computed with the
perturbation theory presented above. The points are consequents obtained by
integrating the full equations of motion from one phase-space point on each
full curve. The orbit shown in Fig.~\ref{fig:overplot} produces consequents
that lie just outside the outermost of the curves of trapped orbits.
The
agreement between the points and the full curves is excellent, both for the
four orbits that librate and for the fifth, circulating orbit, on which
$I=1.05I_{\rm bot}$.

The lower panel of Fig.~\ref{fig:sos2} is similar to the upper panel but for
the case that the perfectly  resonant torus has $J_r=0.035\kpc^2\Myr^{-1}$.
The orbit with the maximum amplitude of libration around this resonant torus comes close
to the torus $J_r=0$ in the course of its libration, so the innermost blue
curve is small.

To obtain the results from perturbation theory plotted in
Fig.~\ref{fig:sos2}, one has to be able to predict $(\vx,\vv)$ given
arbitrary values $(\vtheta,\vJ)$. Given a value of $\vJ$, the associated
torus is constructed by interpolation on a grid of ten tori that lie along
the edge of a rectangle five grid points wide and two grid points high in the
$(J_1',J_3')$ plane. It proves advantageous for the grid to be uniformly
spaced in $\surd J_r$ rather than in $J_r$ as the worst interpolation errors
arise near the ``skinny'' torus that lies on the $J_\phi$ axis in
Fig.~\ref{fig:Hbar}. The axisymmetric tori that generated the broken blue curves in
Fig.~\ref{fig:sos2} lie along the long axis of the rectangle and were
obtained by interpolating between pairs of tori that have the same value of
$J_1'$. Their near-uniform increase in radius reflects their uniform spacing
in $\surd J_r$. 

The area enclosed by each broken curve in Fig.~\ref{fig:sos2}
is proportional to the value of $J_r$ on the corresponding torus. Thus in the
upper panel all these tori have non-negligible values of $J_r$, but in the
lower panel, associated with a resonant orbit that has less radial action by
a factor $\sim3$, the innermost broken blue curve has a very small area and
thus corresponds to an essentially circular orbit.  The area inside each full
black curve is proportional to the value of the resonant action $\cJ$ (eqn
\ref{eq:resJ}).  At the centre of the island of full curves would lie the
single consequent generated by the closed trapped orbit $\cJ=0$, which we
can obtain by setting $I=I_{\rm top}$.

\subsection{Trapping at corotation}\label{sec:CR}

Unfortunately, a
slight adjustment to the theory of Section~\ref{sec:ptheory} is required
before we apply it to the corotation resonance because at corotation the slow
angle is $\vN\cdot\vtheta=\theta_\phi$, so if equations~(\ref{eq:defsJp}) are
to apply we must re-order our unperturbed angles and actions such that
\begin{eqnarray}
(\theta_1,J_1)&=&(\theta_\phi,J_\phi)\nonumber\\
(\theta_3,J_3)&=&(\theta_r,J_r).
\end{eqnarray}
 This done, the previous analysis is valid with $\vN=(1,0,0)$, with the
result that we do not need to distinguish between $\vJ$ and $\vJ'$. The
impact of the non-resonant terms is now as follows. A term changes the
actions by
\begin{eqnarray}
\delta J_r&=&\int\d t\,\dot J_r=\int {\d\theta_r\over\Omega_r}\dot
J_r\nonumber\\
&=&-{2h_{20n}\over\Omega_r}\cos(n\theta_r+2\theta_\phi+\psi_{20n})\\
\delta J_\phi&=&\int\d t\,\dot J_\phi=\int{\d\theta_r\over\Omega_r}\dot
J_\phi\nonumber\\
&=&-{4h_{20n}\over n\Omega_r}\cos(n\theta_r+2\theta_\phi+\psi_{20n}).\nonumber
\end{eqnarray}

\begin{figure}
\centerline{\includegraphics[width=.8\hsize]{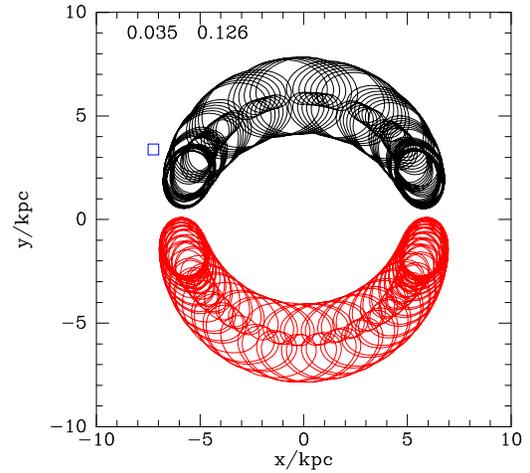}}
 \caption{An orbit trapped at corotation plotted in black by integration of
 the full equations of motion and  in red using perturbation
theory after reflection in the $x$ axis.  The
orbit, which reaches $z=270\pc$, has $J_r=0.035\kpc^2\Myr^{-1}$ and action of
libration is $\cJ=0.184\kpc^2\Myr^{-1}$. The blue square shows the likely
location of the Sun.}\label{fig:corot}
\end{figure}

\begin{figure}
\includegraphics[width=\hsize]{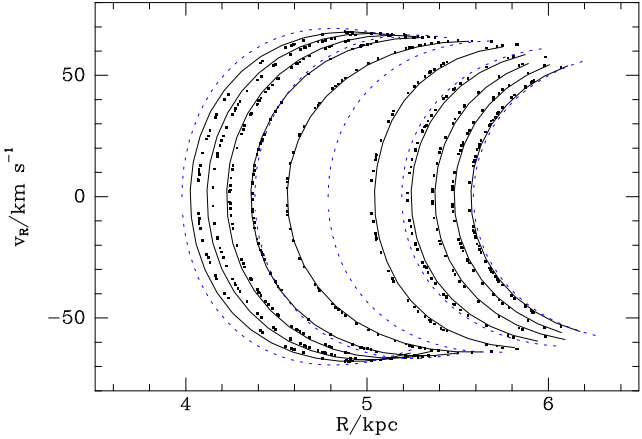} \caption{Surface of section
$\phi=\pi/2$, $z=0$ at $J_r=0.035\kpc^2\Myr^{-1}$,
$J_z=0.0025\kpc^2\Myr^{-1}$. The full black curves and the points are for five
orbits trapped at corotation, the curves being computed with perturbation
theory and the points being consequents from numerically integrated orbits
launched from one point on each curve. The actions of libration are
$\cJ=0.140,0.098,0.066,0.037,0.011\kpc^2\Myr^{-1}$. The broken blue curves
are computed from the three tori of $H_0$ that were used as the basis of
interpolation when computing the full black curves.}\label{fig:corotSOS}
\end{figure}

The resonant term in the Hamiltonian $h_{200}\cos(2\phi+\psi)$ has
$\psi=\pi$, so trapped orbits librate around $\theta_\phi=\pi/2$, which
corresponds to a point on the minor axis of the potential -- in the limit
$J_r\to0$ this coincides with the Lagrange point L$_4$. Fig.~\ref{fig:corot}
shows a trapped orbit, in black as computed by integration of the full
equations of motion and in red as computed from perturbation theory after
reflection in the $x$ axis to avoid overplotting the numerically integrated
version.  This orbit has a relatively large amplitude of libration so it
reaches all the way to the long axis of the bar. In fact, this orbit supports
the bar quite strongly because the star lingers at the turning points of its
libration, when it lies near the $x$ axis. The loops along the orbit's inner
and outer crescents are clearly arise from radial oscillations that have a
much shorter period than the period of libration. The only significant
difference between the red and black versions of the orbit is a small change
in the ratio of these periods.

Fig.~\ref{fig:corotSOS} is a surface of section $\phi=\pi/2, z=0, \dot\phi>0$
for orbits with the same radial and vertical actions
[$(J_r,J_z)=(0.035,0.0025)\kpc^2\Myr^{-1}$] as the orbit shown in
Fig.~\ref{fig:corot}. The full curves are the predictions of perturbation
theory for five values of $I$ while the data points were generated by
integrating the full equations of motion from a single point on each of the
perturbed tori. Each orbit generates two crescents: the orbit with the
largest libration amplitude generates the crescents on the extreme left and
the extreme right, the orbit with the second largest libration amplitude
generates the next two crescents in, and so on. As is to be expected, the
agreement between the data points and the full curves deteriorates as the
amplitude of libration increases.

Since we have imposed the requirement $\dot\phi>0$, all points arise from
times at which the star lies inside its guiding-centre radius and thus is
moving progradely in the corotating frame.  That is, all points arise not too
far from pericentre. If the restriction $\dot\phi>0$ is lifted, each crescent
is complemented by an oppositely directed crescent so each orbit contributes
two ellipses to the diagram, one centred on smaller radii than the other. The
figure is then too busy to be helpful.

The broken blue curves in Fig.~\ref{fig:corotSOS} show the points
generated by three tori of $H_0$ constructed by interpolation along
the centre of the $3\times2$ grid of tori from \TM\ that underpins the
construction of the non-axisymmetric tori. Both the
data points and the full curves track these broken curves to a remarkable
extent. However, in an axisymmetric potential each orbit would generate only
{\it one} crescent, that associated with its angular momentum. In the
non-axisymmetric potential orbits generate two crescents because they cross
the minor axis with one of two values of $J_\phi$. The alignment of the full
and broken curves signifies that the angular momentum with which a star returns
to the minor axis is independent of the phase of its radial oscillations.

\begin{figure}
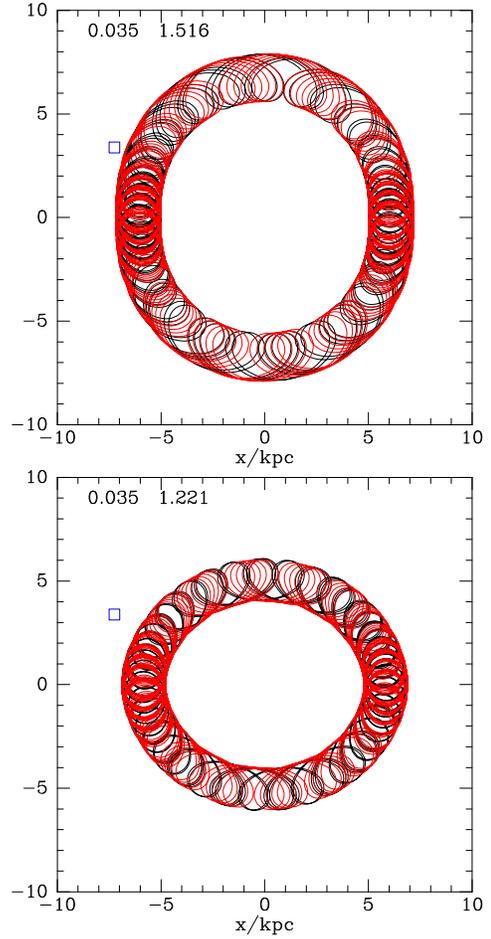

\centerline{\includegraphics[width=.8\hsize]{figs/circulate.ps}}
\centerline{\includegraphics[width=.8\hsize]{figs/circulate_i.ps}}
\caption{Orbits that circulate just outside (top) and inside (bottom)
corotation. The colour scheme is that of
Fig.~\ref{fig:corot}. The upper orbit has extended angular-momentum action
$J_\phi=1.52\kpc^2\Myr^{-1}$, while the lower orbit has
$J_\phi=1.22\kpc^2\Myr^{-1}$. Both orbits have
$(J_r,J_z)=(0.035,0.0025)\kpc^2\Myr^{-1}$.}\label{fig:circulate}
\end{figure}

\begin{figure}
\includegraphics[width=\hsize]{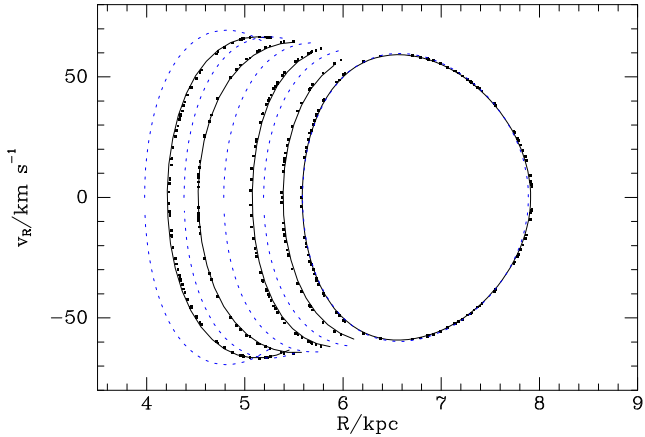}
 \caption{In this surface of section $\phi=\pi/2,z=0$ the circulating upper
orbit of Fig.~\ref{fig:circulate} generates the complete circuit of
consequents on the right. The full curve was
computed from perturbation theory. Also shown on the left are the consequents
and invariant curves of two librating orbits that differ from the circulating
orbit in having $I>I_{\rm bot}$. The broken blue curves show the cross
sections of tori produced by \TM\ that provided the basis for perturbation
theory.}\label{fig:circulateSOS}
\end{figure}

\subsubsection{Circulating orbits}

When \TM\ fits tori to the full Hamiltonian, as in B16, everything one needs
to know about an orbit that circulates rather than librates can be obtained
directly from \TM. Here, however, \TM\ fits tori to only the axisymmetric
part of the Hamiltonian, so perturbation theory is required to obtain a good
fit to an orbit that circulates -- in fact, the
orbit
can be significantly non-axisymmetric. Fig.~\ref{fig:circulate}
illustrates this point by showing orbits that have slightly too much (upper panel)
and slightly too little (lower panel) angular momentum to be trapped by the
corotation resonance.  Both orbits are far from axisymmetric.  As in
Fig.~\ref{fig:corot} the black curves show the orbits
obtained by direct integration of the equations of motion and the red curves
show the orbits obtained with perturbation theory. The main difference
between the black and red curves is the spacing between the loops to which
radial oscillations give rise: along the black curve in the upper panel the
loops are closer together, indicating that $\Omega_r/\Omega_\phi$ lies close
to an integer, whereas along the red curve the loops are more clearly
separated indicating a value of $\Omega_r/\Omega_\phi$ that lies less close
to an integer. 

\begin{figure}
\centerline{\includegraphics[width=.8\hsize]{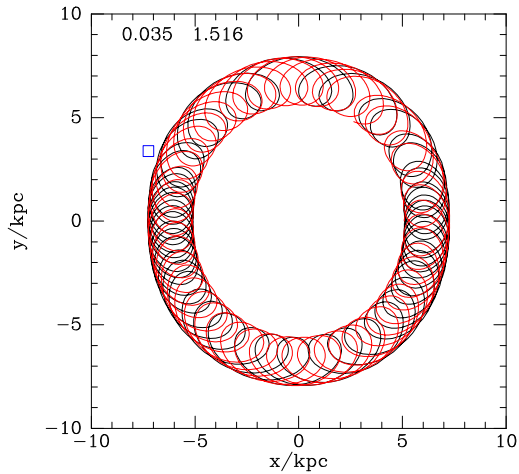}}
\caption{The orbit shown in the upper panel of Fig.~\ref{fig:circulate}
computed using non-resonant perturbation theory. The general shape of the
orbit is correctly reproduced but the orbit fails to linger near the
$x$ axis.}\label{fig:nores_circ}
\end{figure}

Fig.~\ref{fig:circulateSOS} shows a surface of section.  The orbit
plotted in the upper panel of Fig.~\ref{fig:circulate} produces the complete
circuit of consequents on the right. The curve on which they lie was computed
with perturbation theory. To set
the context, we show on the left of this surface of section the consequents
and invariant curves of two orbits with the same values of $J_r$ and $J_z$
but with $I>I_{\rm bot}$, so they librate. For these orbits, as in
Fig.~\ref{fig:corotSOS}, we only show consequents associated with
$\dot\phi>\omegap$, whereas for the circulating orbit we show consequents for
both signs of $\dot\phi-\omegap$.  This is why the invariant curve of the
circulating orbit closes on itself, whereas the invariant curves of the other
two orbits form pairs of crescents, one for each of the two values of angular
momentum at which the star crosses the bar's minor axis. The broken blue
curves show cross sections through five axisymmetric tori located along
the central spine of the $5\times2$ grid of tori from \TM\ that formed the
basis for interpolation. The broken blue curve on the right, associated with the
largest angular momentum, keeps quite close to the black invariant curve of
the circulating orbit, indicating that as the latter passes the minor axis,
its angular momentum lies in a narrow range. This does not imply, however,
that the orbit's angular momentum is nearly constant: near the major axis it
has substantially more angular momentum.

The above reconstruction of a circulating orbit was obtained from the
pendulum equation. As such it relies on the determination of the Fourier
coefficients $h_n$ on the perfectly resonant torus, which is at some distance
from it in phase space. 

As was noted in Section \ref{sec:resornonres}, when an orbit is obtained from
resonant perturbation theory, one is using  Fourier coefficients of $H$ that
are becoming increasingly stale as one progresses away from the region of
entrapment. At some point  it becomes
expedient to adopt non-resonant perturbation theory, for then the Fourier
coefficients employed are computed locally, on the unperturbed tori over
which the orbit is actually ranging. The analytic work of \cite{MonariDF2016}
is based on this type of perturbation theory. 

The red curve in Fig.~\ref{fig:nores_circ} shows the orbit that circulates
just outside the region of entrapment by corotation as obtained from
non-resonant perturbation theory, with the black curve as usual showing the
result of direct numerical integration. We see that non-resonant perturbation
theory recovers the shape of the orbit well, but comparison with the upper
panel of Fig.~\ref{fig:circulate}, which shows the same orbit computed with
resonant perturbation theory, reveals a failure to recover the way the orbit
lingers along the $x$ axis. This is a natural consequence of the key
approximation of non-resonant theory, namely that the unperturbed angles
evolve uniformly in time. Further from the boundary of the region of
entrapment, the non-uniform evolution of the unperturbed angle variables
becomes less marked and non-resonant perturbation theory comes into its own. 

In conclusion, resonant perturbation should be used both inside and close to
the region of entrapment, while non-resonant theory should be used elsewhere.

\subsection{Trapping at the ILR}\label{sec:ILR}

To obtain orbits trapped at the inner Lindblad resonance, the code described
in Appendix~\ref{sec:TMclass} should be used to construct an instance of {\tt
resTorus\_L} with the resonant vector set to $\vN=(1,0,-2)$. In contrast to
the CR and OLR, at the ILR $G$ proves positive while $\psi_\vN$ vanishes.  As a
consequence, libration is still around $\theta_1'=\pi$ but now the torus with
$I=I_{\rm bot}$ has vanishing libration amplitude while maximum libration
amplitude is attained at $I=I_{\rm top}>I_{\rm bot}$. Values of $I$ smaller
than $I_{\rm bot}$ are now forbidden, while values larger than $I_{\rm top}$
correspond to circulating orbits. 

Examples of orbits trapped at the ILR will
be given in a forthcoming publication that focuses on  resolving difficulties
that the current
version of \TM\ encounters with orbits that have small $J_r$ and large $\cJ$.

\section{Structure of local velocity space}\label{sec:Vspace}

Now we use our non-axisymmetric tori to examine the structure of the velocity
distribution of solar-neighbourhood stars. This exercise illustrates both the
inconvenience of orbit-based models of galaxies and the relative strength of
torus modelling among orbit-based techniques. We define $U=v_{R\odot}-v_R$
and $V=v_\phi-v_{\phi\odot}$, so the Sun is at $(0,0)$, $U$ increases with
motion towards the Galactic centre and $V$ is the amount by which the star's
motion in the direction of Galactic rotation exceeds that of the Sun. We
assume that $(v_{R\odot},v_{\phi\odot})=(-11,\Vc+12)\kms$ \citep{SBD2010}. For
reasons that will become apparent, most results are presented for pattern
speed $\omegap=0.038\Myr^{-1}$.

\subsection{Orbit-based modelling}

When one wishes to compute the density or the velocity distribution at a
given point in a model, it is enormously desirable to be in possession of an
algorithm that computes $(\vtheta,\vJ)$ given $(\vx,\vv)$, for then the
density at position $\vv$ in velocity space is simply $f(\vJ)$ and the
density can be recovered by integrating over $\vv$. For an axisymmetric
model, or a triaxial model that has zero pattern speed, we can use the St\"ackel
Fudge \citep{JJB12:Stackel,SaJJB15:Triaxial} for this purpose, but the Fudge
has yet to be generalised to systems with rotating non-axisymmetric
potentials. Algorithms for non-axisymmetric systems that have been used with
some success are those of \cite{WD99:Bar} and \cite{MonariDF2016}.
\cite{WD99:Bar} evaluated $f(\vx,\vv)$ by integrating an orbit from
$(\vx,\vv)$ backwards in time while the bar amplitude was reduced to zero,
and then evaluated $f$ at the computed location in the axisymmetric model.
\cite{MonariDF2016} computed $f(\vx,\vv)$ in essentially the same way except
they followed orbits backwards analytically using non-resonant perturbation
theory in axisymmetric angle-action coordinates. These algorithms rely on the
adiabatic invariance of $\vJ$ and consequently fail when resonant trapping is
important.

Given the lack of robust algorithms for computing $(\vtheta,\vJ)$ from
$(\vx,\vv)$ in rotating, non-axisymmetric  potentials, nearly all modelling
of non-axisymmetric systems (and much modelling of axisymmetric ones too) has
been orbit based in the sense that one finds orbits and assigns them weights
such that available observational data are consistent with the values
predicted by the weighted averages of the orbits. The classic technique is
that of \cite{Sc79} but important advances have recently made
with the ``Made-to-Measure'' (M2M) technique proposed by \cite{SyerTremaine} and
developed by \cite{deLorenzi2007}, \cite{DehnenM2M}, \cite{Long2010} and \cite{MorgantiGerhard} among others.

The Schwarzschild and M2M techniques differ in (i) how weights are assigned,
and (ii) how orbits are used: for each orbit M2M integrates orbits in
parallel and for each orbit holds $(\vx,\vv)$ for only a single time, whereas
Schwarzschild integrates orbits in series and for each orbit retains
$(\vx,\vv)$ at a large number of times. Hence Schwarzschild uses orbits as
time series, and computes densities by summing the time intervals during
which each star is in a small volume around the point at which the density is
required. If orbits are replaced by tori, it becomes easier to compute the
density because one then has analytic formulae for $\vx(\vtheta)$ and
$\vv(\vtheta)$ so one can solve for the angles (if any) $\vtheta$ at which the torus
visits a given location and evaluate the velocities with which the visits
occur. This approach underlies the code for chemodynamical evolution used by
\cite{SchoenPJM}, and also the analysis of data from the LAMOST survey
\citep{LAMOST} presented by \cite{LAMOSTmodel}. Our non-axisymmetric,
sometimes resonantly trapped, tori allow us the extend this technique to the
non-axisymmetric problem posed by the solar neighbourhood.

\begin{figure}
\includegraphics[width=\hsize]{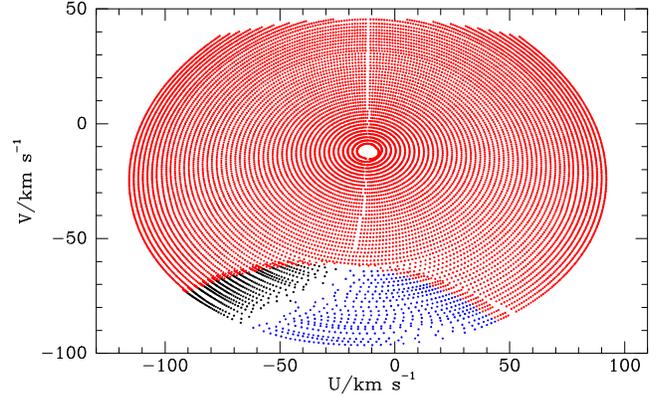} \caption{Each dot shows a
velocity $(U,V)$ at which a torus visits the Sun-like location
$(R,z,\phi)=(8\kpc,10\pc,155^\circ)$. Points contributed by circulating tori
are plotted in red while points contributed by tori trapped at corotation are
plotted in blue/black. The bar's pattern speed is $0.038\Myr^{-1}$, 6850
tori contribute to the plot, and on the outermost red ellipse
$J_r=0.12\kpc^2\Myr^{-1}$.}\label{fig:UVdot}
\end{figure}

The general plan is to compute tori for a grid in action space and find the
velocities (if any) at which each torus visits the Sun. Then the density of
the corresponding cells in the $UV$ plane is incremented by
\[\label{eq:UVincr}
{f(\vJ)\over f_{\rm s}(\vJ)}{\p(\vv)\over\p(\vJ)},
\]
 where $f_{\rm s}$ is the sampling density that is defined by the grid.  The
Jacobian, which converts from density in action space to
density in velocity space, is provided by \TM. Indeed with $(\vx,\vv)$
denoting Cartesian phase-space coordinates and $S$ the generating function of
the canonical transformation to angle-action coordinates,
\[
\vtheta={\p S\over\p\vJ}\hbox{ and }
\vv={\p S\over\p\vx},
\]
so
\[
{\p\vv\over\p\vJ}\bigg|_\vx={\p^2S\over\p\vJ\p\vx}={\p\vtheta\over\p\vx}\bigg|_\vJ.
\]
Taking determinants it follows that
\[
{\p(\vv)\over\p(\vJ)}=1\bigg/{\p(\vx)\over\p(\vtheta)}.
\]
\TM\ does not directly provide the Jacobian on the right because it uses
cylindrical polar coordinates. So it provides
\[
{\p(R,z,\phi)\over\p(\vtheta)}={\p(\vx)\over\p(\vtheta)}{\p(R,z,\phi)\over\p(\vx)}
={\p(\vx)\over\p(\vtheta)}{1\over R}.
\]

\subsection{Orbits trapped at corotation}

Each point in Fig.~\ref{fig:UVdot} shows the velocity with which a torus
visits the Sun-like location $R_0=8\kpc$ $\phi=155\,$deg. We plot in red
visits by tori that are not resonantly trapped, and in blue or black visits
by tori that have been trapped by corotation. If a torus reaches the Sun, it
usually does so at four distinct velocities, two with $v_z>0$ and two with
$v_z<0$. In Fig.~\ref{fig:UVdot} we show only visits with $v_z>0$. Then the
plot should contain two visits per torus, one on the left and one on the
right of a roughly horizontal line. In the region of entrapment at the bottom
of the figure, the left-hand visits have been coloured black and the
right-hand ones coloured blue. As the action of libration $\cJ$ diminishes,
the left- and right-hand points converge on the rather bare space where the
black and blue points approach one another.  This bare region arises because
the tori have been sampled uniformly in $I$: the region can be eliminated by
sampling more densely at small values of $|I|$.

\begin{figure}
\includegraphics[width=\hsize]{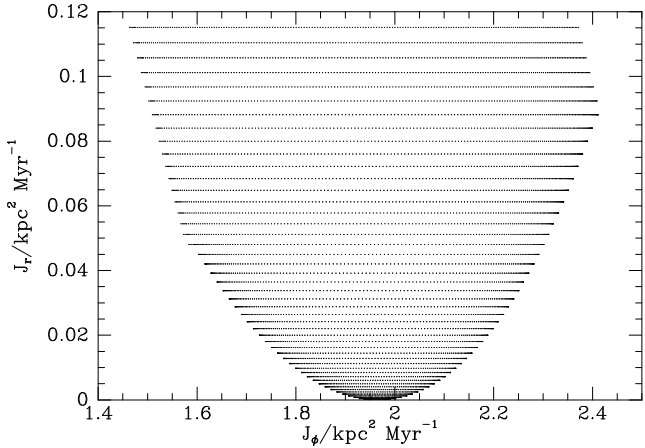}
\caption{The distribution in action space of the tori that generated the red
dots in Fig.~\ref{fig:UVdot}. The parabolic shape at the base is defined by
the requirement to be at apo- or peri-centre at the Sun. The clip off the top right corner
is to avoid orbits trapped by the OLR. Similarly, the upper left edge is
defined by avoidance of orbits trapped by the CR. Note the increased density
of dots at the ends of each horizontal line and the increasing density of
lines towards the base.}\label{fig:Jspace}
\end{figure}

The dots in Fig.~\ref{fig:UVdot} are arranged in rings because the
tori were computed on a grid in action space: for $0\le n< N$, the grid in
$J_r$ is defined by
\[
J_{r,n}=\left[\sqrt{J_{r\,\rm min}}+{n\over N}\sqrt{J_{r\,\rm max}}\right]^2
\]
with $J_{r\,\rm min}=0.0002\kpc^2\kms$, $J_{r\,\rm max}=0.12\kpc^2\kms$ and
$N=50$. Fig.~\ref{fig:Jspace} shows the grid points in action space. By
making the grid points uniform in $\surd J_r$ we ensure a near uniform
distribution of rings in velocity space. Choosing grid points in $J_\phi$ is
challenging because one wants to obtain points near the line $U=v_{R\odot}$.
In the lower part of Fig.~\ref{fig:UVdot} and in an axisymmetric system,
these points would correspond to visits of stars with low $J_\phi$ that reach
the Sun at apocentre, while in the upper portion of the diagram the points
correspond to visits at percentre by stars with large $J_\phi$. In a
non-axisymmetric potential the situation is more complex, but the general
idea holds. So for given $J_r$ we compute the values of the generalised
angular momentum $J_\phi$ of visits at apocentre ($J_{\rm apo}$) and at
pericentre ($J_{\rm peri}$) and distribute our grid points between these
lower and upper bounds on $J_\phi$. We need grid points closely spaced near
the limiting values because as $U\to v_{R\odot}$ the Jacobian in equation
(\ref{eq:UVincr}) diverges.  Specifically we take
\[\label{eq:JphiGrid}
J_{\phi,k}=\fracj12\bigl[J_{\rm peri}+J_{\rm apo}-(J_{\rm peri}-J_{\rm
apo})\cos((k+\fracj12)\pi/K)\bigr]
\]
 for $0\le k<K$ in integer step.  Then the divergence in the
Jacobian is cancelled by a matching divergence in the sampling density
$f_{\rm s}$ that reflects divergence of the density of grid points as
$U\to v_{R\odot}$, which is just about visible in Fig.~\ref{fig:Jspace}.

Fig.~\ref{fig:UVdot} reveals that this strategy has not been entirely
successful in that a river of white can be discerned running vertically
through most of the red region. In addition, at the very centre of the figure
there is a white patch reflecting the finite smallest value
$J_r=0.0002\kpc^2\Myr$ adopted: \TM\ has difficulty computing tori for very
small actions. This hole could be filled with tori computed with the epicycle
approximation, which is at its most precise in this region.

It is worth noting that the difficulty sampling the $UV$ plane at small $|U|$
is directly related to the fact that the real-space density generated by an
orbit always diverges near its boundaries, and by Liouville's theorem, its
velocity-space density has to be low there. Put another way, the
velocity-space density along $U= v_{R\odot}$ is inevitably contributed by a small
number of stars that each contribute a large probability density at $U= v_{R\odot}$.

In Fig.~\ref{fig:UVdot} another blemish is evident at about four o'clock: a
wedge that is clear of red dots. The bulk of the red points are computed
using non-resonant perturbation theory, while the points that bound the
region of trapped orbits are computed with resonant perturbation theory --
red points are contributed by circulating orbits and black/blue points by
librating orbits. Since resonant and non-resonant perturbation theory involve
different approximations, their results do not match perfectly across the
join (cf Figs~\ref{fig:circulate} and \ref{fig:nores_circ}). In particular,
at the left-hand edge of the red crescent, similar velocities are obtained
with different values of $J_\phi$, and when this happens, the figure is
redder than usual. At the right-hand edge of the crescent, this
over-population is compensated by a small empty region. This blemish might be
smoothed by shifting the transfer to non-resonant theory further from the
region of entrapment.

\begin{figure}
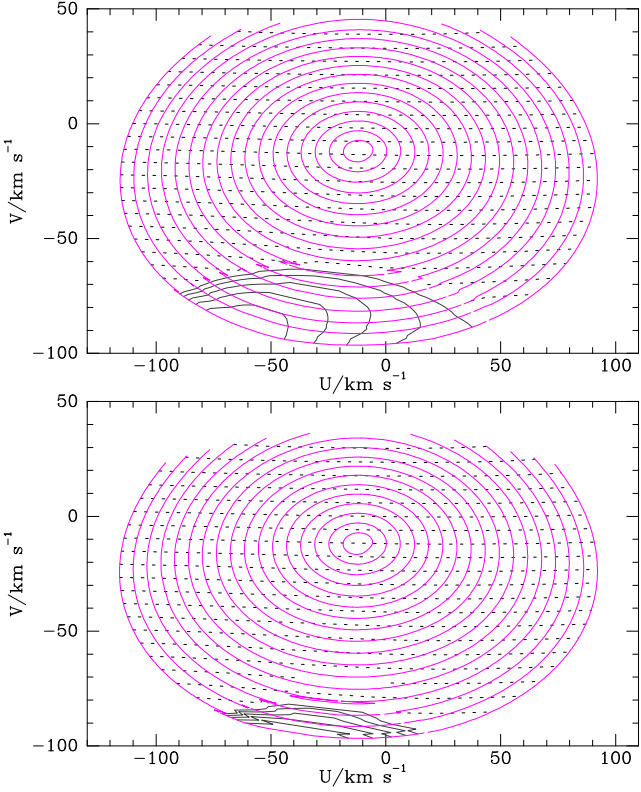

\includegraphics[width=\hsize]{figs/UVcont38.ps}
\includegraphics[width=\hsize]{figs/UVcont40.ps}
\caption{Contours in the $UV$ plane on  of constant integrals of motion.
$J_r$ is constant on the
magenta curves, while generalised angular momentum, $J_\phi$, is constant on
the dotted curves. The full grey curves are contours of constant action of
libration around CR, $\cJ$. The upper panel is for $\omegap=0.038\Myr^{-1}$, while the
lower panel is for  $\omegap=0.040\Myr^{-1}$.}\label{fig:UVconts}
\end{figure}

Each point in Fig.~\ref{fig:UVdot} is associated with a value of $J_r$ and a
value of either $J_\phi$ (if circulating) or $\cJ$ (if librating).
Fig.~\ref{fig:UVconts} shows the curves (magenta for $J_r$, broken blue for
$J_\phi$, black for $\cJ$) on which these integrals are constant. All points
share the same value $J_z=0.0025\kpc^2\kms$. In the absence of the bar's
contribution to the potential $\Phi_2$, we would have
$J_\phi=R_0(v_{\phi\odot}+V)$ so contours of constant $J_\phi$ would be
straight horizontal lines. $\Phi_2$ modifies $J_\phi$ such that its contours
(broken blue) cease to be horizontal and straight.  In the region of trapping
by CR, $J_r$ (magenta) is complemented by $\cJ$, the action of libration,
whose (grey) contours form asymmetric arches. 

\subsection{Orbits trapped at the OLR}

The upper panel of Fig.~\ref{fig:UVconts} is for pattern speed
$\omegap=0.038\Myr^{-1}$, while the lower panel is for
$\omegap=0.040\Myr^{-1}$.  Comparison of these panels reveals how sensitive
local velocity space is to the pattern speed. This sensitivity is a
consequence of how close the Sun is to the edge at $R\sim R_{\rm edge}$ of
the region of entrapment by corotation: when $\omegap$ is lowered this edge
shifts outwards and displacement by a distance that is small compared to
$R_0$ is not small compared to $R_0-R_{\rm edge}$.

\begin{figure}
\includegraphics[width=\hsize]{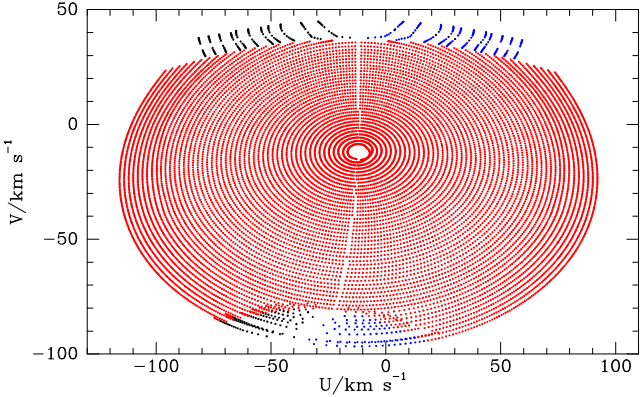}
\caption{As Fig.~\ref{fig:UVdot} but for pattern speed
$\omegap=0.040\Myr^{-1}$ and with the velocities of orbits trapped at the OLR
appearing as `hair' on top. }\label{fig:UVdot40}
\end{figure}

At the top of Fig.~\ref{fig:UVdot}, the curves of constant $J_r$ terminate
where they encounter the region of entrapment by the OLR.
Fig.~\ref{fig:UVdot40} is the corresponding plot for the faster pattern speed
$\omegap=0.040\Myr^{-1}$.  The regions within which orbits are trapped have
both moved down, so the region within which orbits are trapped at OLR is now
clearly visible at the top of the figure.  The black dots generated by orbits
trapped by the OLR line up along curves of constant $J_3'=J_\phi-2J_r$, which
have both left and right branches. Each orbit appears at four points along
the curve associated with its value of $J_3'$. Orbits with large values of
the libration action
$\cJ$ appear towards the top and bottom ends of each stubby line of dots,
while orbits with small values of $\cJ$ appear either side of the middles
of these lines. Hence contours of constant $\cJ$ delineate a highly
elongated and slightly
curved  island. The OLR region in Fig.~\ref{fig:UVdot40}
contains eleven black and eleven blue lines, and each pair of lines corresponds to
the outermost eleven red ellipses, being the ellipses with the largest values of
$J_r$.

\begin{figure}
\includegraphics[width=\hsize]{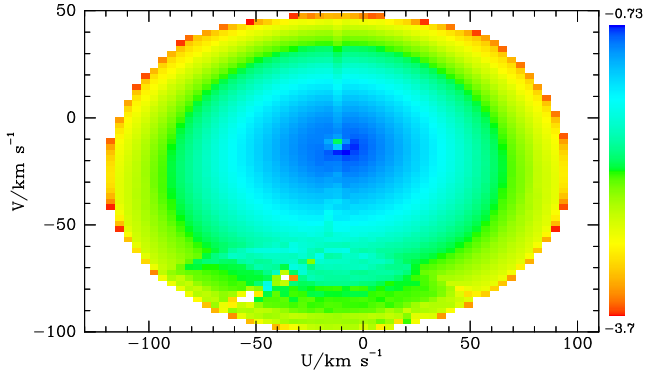}
\caption{A possible density of stars in the $UV$ plane at the Sun-like location
$(R,z,\phi)=(8\kpc,10\pc,155^\circ)$. In the region of circulation, the
density is given by the DF of Piffl et al.\ (2014). In the region
of entrapment, the density is computed from the same DF but using the actions
of the perfectly resonant, axisymmetric orbit regardless of the action of
libration around this orbit. The pattern speed is
$\omegap=0.038\Myr^{-1}$}\label{fig:UVplane}
\end{figure}

\begin{figure}
\includegraphics[width=\hsize]{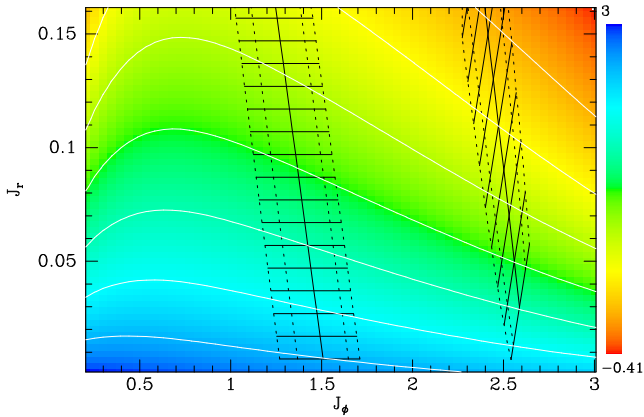} \caption{The colours and
white contours indicate on a log scale the value of the DF for the adopted
axisymmetric Galaxy model. The regions within which orbits are trapped by the
bar with $\omegap=0.038\Myr^{-1}$ at either corotation or the OLR are marked
by ladders as in Fig.~\ref{fig:Hbar}.}\label{fig:DF}
\end{figure}

Fig.~\ref{fig:UVplane} shows the density of stars in the $UV$ plane that is
predicted by a plausible DF $f(\vJ)$.  Colours indicate logarithms
to base 10 of the density. We base our DF on the work of
\cite{Piea14}, who fitted a DF to data from the Sloan Digital Sky Survey
\citep[][SDSS]{Juea08} and the Radial Velocity Experiment
\citep[][RAVE]{RAVE1}. Their DF is based on the assumption that the Galaxy is
axisymmetric and has to be adapted to encompass non-axisymmetry. The colours
and contours in Fig.~\ref{fig:DF} show the value of the DF in the same slice
$J_z=0.0025\kpc^2\Myr^{-1}$ through action space for which
Fig.~\ref{fig:Hbar} shows values of $H_0$. The value of the DF
diminishes rapidly with increasing $J_r$ and more gradually with increasing
$J_\phi$. Since stars trapped at corotation oscillate along the rungs of the
figure's left-hand ladder, they move between regions of high $f$ on the left
and low $f$ on the right. By Jeans' theorem the value of the true DF $f_{\rm
true}$ must be a function of the constants of motion. Consequently, in the
region of trapping we must replace $J_\phi$ in the argument list of $f_{\rm
true}$ by $\cJ$, the libration action, which controls how far each side of
the line of exact resonance a star moves as it librates. In particular, at a
given value of $J_r$, $f_{\rm true}$ must take the same value just to the
right of the dashed line that marks the left-hand boundary of the trapping
region as it does just to the left of the dashed line that forms the
right-hand boundary of the trapping region. In fact, the simplest possible
structure for $f_{\rm true}$ is that it is constant along each of the rungs
of the ladder that marks the trapping region in Figs.~\ref{fig:Hbar} and
\ref{fig:DF}. \cite{Monari2017} adopted this ansatz. The actual population
will be determined by the precise history of bar formation and lies beyond
the scope of this paper. It is, however, a question that can be addressed by
an extension of the modelling technique of \cite{Aumer2016a}.

In the region populated by circulating orbits, we simply evaluate the
\cite{Piea14} DF on the given numbers $(J_r,J_z,J_\phi)$. In the region
populated by trapped orbits, we evaluate the Piffl et al.\ DF using the
actions of the perfectly resonant orbit from which it obtained by adding some
amplitude of libration.
This choice leaves the star density in the  region of entrapment higher than
at points outside this region that are  similar distances from the centre of
the figure. However, within the trapped zone
one can in principle take the DF to be {\it any} non-negative function of the
integrals $(J_r,J_z,\cJ)$.

The observed distribution of local stars in the $UV$ plane shows an
overdensity centred on $\sim(-35,-50)\kms$ that is called the Hercules
Stream.  \cite{WD99:Bar} suggested that it is caused by stars being trapped
by the bar's OLR, and by modelling this process derived $\omega_{\rm
p}=54\kms\kpc^{-1}$.  As was noted in Section~\ref{sec:nonaxi}, there is now
strong evidence that the bar's pattern speed is not greater than $\omega_{\rm
p}=40\kms$, with the consequence that the OLR lies well outside $R_0$ and
could only affect local velocity space at positive values of $V$. With the
lower pattern speed we instead lie not far outside the bar's CR, and
\cite{Perez2017} have shown that the model of the bar developed by
\cite{Wegg2015} with $\omega_{\rm p}=39\kms\kpc^{-1}$ predicts a feature
rather like the Hercules stream that is caused by stars trapped into
libration around the Lagrange point L$_4$. Fig.~\ref{fig:UVplane} shows that
in our model of the bar stars trapped at corotation  appear at significantly
smaller values of $V$: the region of entrapment does not extend above
$V=-62\kms$. On the other hand, it is centred on $U\sim-30\kms$, rather as in
the observations.

The region of entrapment can be moved further up by lowering $\omegap$, or by
adopting a locally rising circular-speed curve.

\section{Discussion}\label{sec:discuss}

\subsection{Relation to previous work}

We have addressed much the same problem as \cite{MonariDF2016} and
\cite{MonariStayAway2017} from a rather different perspective.
\cite{MonariDF2016} use angle-action variables to compute the first-order
change $f_1$ in an unperturbed DF $f_0(\vJ)$ when a non-axisymmetric
component is added to the potential. As a resonance is approached, $f_1$
diverges, so there is a band in action space around a resonant torus within
which $f_0+f_1$ is not non-negative and hence is manifestly in error.  They
show that this problem is not mitigated by going to second order.
\cite{MonariStayAway2017} use the standard pendulum equation to determine the
width of the trapping region, but they do not compute actions for trapped
orbits, and whenever $|f_1|>f_0$ they use
\[
f=\begin{cases}
2f_0&\hbox{if }f_1>0,\\0&\hbox{otherwise}.
\end{cases}
\]
Using this prescription and the angle-action coordinates provided by the
epicycle approximation, \cite{MonariStayAway2017} compute the structure of
the $UV$ plane for a variety of pattern speeds and bar strengths. The main
differences with our work are (a) we use tori rather than the epicycle
approximation so we retain accuracy even for eccentric orbits, (b) we are not
restricted to planar orbits, and (c) we
have complete control of the DF of trapped orbits, so our DF is finite but
discontinuous at the edges of the trapping region as it should be. The
practicalities of the two calculations differ significantly because tori
provide the transformation $(\vtheta,\vJ)\to(\vx,\vv)$ while the epicycle
approximation provides the inverse transformation
$(\vx,\vv)\to(\vtheta,\vJ)$. The latter is in many ways the more convenient
direction, and by using the St\"ackel fudge \citep{JJB12:Stackel} for an
axisymmetric system one can proceed in this direction without losing accuracy
for eccentric orbits.  However, the St\"ackel fudge is not applicable to rotating
non-axisymmetric potentials and it does not tackle resonant trapping, so
currently the impact of resonant trapping can only be reliably computed using
tori. 

Although resonant trapping changes individual orbits qualitatively,
Fig.~\ref{fig:UVplane} demonstrates that the impact of trapping on velocity
space {\it may} be quite modest, depending on the form that the DF takes in
the region of entrapment. This conclusion probably explains why axisymmetric
models such as that of \cite{Piea14} have yielded excellent fits to data.

\subsection{Poisson noise and modelling strategy}

The main difficulty encountered when using tori is getting the sampling
density right. For example, in a plot of the $UV$ plane one wants a dense
sampling near the peak of the star distribution, that is near the local
standard of rest. Here $U$, $V$ etc, are small and small changes in $U,V,W$
correspond to tiny changes in $J_r\propto U^2$ and $J_z\propto W^2$. Hence
tori must be constructed on a very non-uniform grid in action space. It is
similarly necessary to sample densely in $J_\phi$ near the values for
which a star reaches apo- or peri-centre at the location for which the
$UV$ plane is required.  Fortunately, torus interpolation makes dense
sampling in a particular region computationally affordable, and
Fig.~\ref{fig:UVplane} demonstrates that the non-uniform action-spaced grid
of Fig.~\ref{fig:Jspace} provides adequate sampling of
circulating orbits. The sampling of resonantly trapped orbits needs
refinement, however.

Custom sampling of action space can be avoided by devising an algorithm that
interpolates to a general point in velocity space the values taken on a
sub-optimal grid by the constants of motion, $(J_r,J_z,J_\phi)$ or
$(J_r,J_z,\cJ)$ as appropriate. An algorithm of this type was used to plot
the contours of Fig.~\ref{fig:UVconts}, and a similar algorithm lies at the
heart of the chemodynamical evolution code that underpins \cite{SchoenPJM}.
This procedure seems to be the most promising way forward for chemodynamical
modellers.
 
Poisson noise is a constant irritant when one seeks to model galaxies with
precision, so it is a major headache when modelling the Milky way. Indeed, to
extract a reasonable plot of the local $UV$ plane from the M2M model of
\cite{Wegg2015}, \cite{Perez2017} had to upgrade  the model significantly
because the solar neighbourhood contains only a very small fraction of the
Galaxy's stars. Plots such as Fig.~\ref{fig:UVplane} show that the
combination of $f(\vJ)$ modelling and torus interpolation enables one to beat
Poisson noise down to low levels at small computational cost: even with the
current not-optimised version of \TM\ this figure can be computed from
scratch in of order half a processor-hour. The $UV$ plane predicted by a
different DF can be computed in less than a second. 

In standard Schwarzschild modelling, one starts from a potential $\Phi(\vx)$
and seeks orbit weights that are consistent with $\Phi$ and observational
constraints.  Unless the system is extremely simple, it is not unlikely that
$\Phi$ cannot be generated by any set of non-negative weights. When tori are
employed, one can start with $\Phi$ and adjust the weights of tori (or the
parameters of DFs for each component) until reasonable agreement with data is
achieved. Then one can relax the potential to exact self-consistency
\citep{JJB14,PifflPenoyreB} to obtain a completely sound model. Standard
Schwarzschild modelling does not permit this step to self-consistency.

In summary, there are many reasons why classical Schwarzschild modelling, in
which orbits are used as time sequences, is abandoned in favour of torus
modelling: tori are simply more compact and very much more powerful than
classical orbits. In point of historical fact, Martin Schwarzschild himself
took the first step towards torus modelling \citep{RatcliffChangSch}.

\subsection{Accuracy and chaos}

The quality of the fits delivered by perturbation theory is remarkable. For
example, the length of black ladder rungs in Fig.~\ref{fig:Hbar} and the
qualitative difference between the full black and broken blue invariant
curves in Fig.~\ref{fig:sos2} show that along an orbit trapped by the OLR the
angular momentum and radial action typically make large excursions and
trapped orbits do not keep close to the perfectly resonant torus which
provides the numbers used to construct a whole family of trapped and even
circulating orbits. Yet surfaces of section and plots in real space
demonstrate that orbits can be accurately reconstructed perturbatively.  On
account of a poor initial choice of the bar strength, most of the figures
presented have been plotted for a bar that is almost twice as strong. The
quality of fits in those figures is not much inferior to that shown here.
Thus the Galaxy does not require us to push perturbation theory near its
limits. It is remarkable that such impressive results can be obtained by
averaging to zero the non-resonant terms in the Fourier expansion of the
Hamiltonian. The legitimacy of this procedure is by no means self-evident.

The extent to which the dynamics of disc stars is independent of $J_z$ is
striking. In particular, stars from the thin and thick discs that have
similar values of $J_r$ and $J_\phi$ will respond to the bar in essentially
identical ways. 
 
Another remarkable feature of this study is the absence of any sign of chaos.
It is very likely that this absence reflects the pure $m=2$ nature of the
applied non-axisymmetry. A natural next step is to add a smaller $m=4$
component, see what additional resonances emerge, and understand if/how these
give rise to some chaotic regions. Numerical integrations of chaotic orbits
suggest that they are piece-wise quasi-periodic so they can be understood as
orbits that move between tori belonging to different resonantly trapped
families. The prospect of bringing order into chaos by representing
orbits as superpositions of tori is an exciting one.

\subsection{Spiral structure}

The technique explored here in the context of the bar is equally applicable
to a steady spiral perturbation. Whether the assumption of a constant pattern
speed is at all useful in the context of spiral structure remains to be seen,
however.  Since the gravitational field of a tightly wound spiral pattern
decays more rapidly with $|z|$ than that of a bar, one would expect the
Fourier decomposition of a spiral arm to have more power in terms with
non-vanishing $k_z$ than the bar provides. Consequently, contrary to what we
have found in the case of the bar, we expect significant differences between
the responses of thin- and thick-disc stars to a spiral.  Moreover, if a
spiral pattern can bring $J_z$ into play, it will lead to more interesting
dynamics than we have encountered from the bar. Finally, realistic spiral
patterns will have significant power in more than one azimuthal quantum
number, and thus raise the prospect of cross-talk between resonances.
Exploring these issues will require significant effort.

\subsection{Connection to gyrokinetics}

Figs.~\ref{fig:corot} and \ref{fig:circulate} are strongly reminiscent of
the gyrokinetic theory of a magnetised plasma. In that theory
one computes the dynamics of the points around which electrons rapidly gyrate
with an adiabatically  conserved magnetic moment. Near CR there are two
adiabatically conserved quantities, $J_r$ and $J_z$. Moreover
Fig.~\ref{fig:corotSOS} indicates that although a star's angular momentum
changes significantly along an orbit trapped by CR, it is a unique
function of the location of the gyrocentre, i.e., $\theta_\phi$. This is
precisely the prediction of perturbation theory to the extent that it is
possible to neglect the non-resonant terms: the angular momentum is
$J_\phi=J_{\phi\,\rm res}+\Delta(\theta_\phi)$. To a good approximation
dynamics near corotation can be reduced to the one-dimensional motion of
quasi particles that are endowed with specified quantities of radial and
vertical action.

\section{Conclusions}\label{sec:conclude}

Ours is a barred Galaxy, so work with axisymmetric Galaxy models is
inherently limited. In fact, it is in many ways surprising what good fits one
can obtain with axisymmetric models to data for
stars in the extended solar neighbourhood \citep{BinneyBurnett}. As the
available data become richer and more precise, we cannot avoid progressing to
employing non-axisymmetric models for the interpretation of astrometric data.

The impact of non-axisymmetric components of the potential is typically
localised to the regions of phase space in which a resonant condition is
nearly satisfied. A key but easily overlooked point here is that orbital tori
are required even to identify these regions -- the concept of resonance is
inextricably tied up with that of angle-action coordinates and thus with
orbital tori. Given an axisymmetric gravitational potential, \TM\ can be used
to construct a system of angle-action coordinates for a very closely related
Hamiltonian $H_0(\vJ)$. Any difference between $H_0(\vJ)$
and the true Hamiltonian is liable to have a significant impact on orbits in
a region of phase space within which a resonant condition is nearly
satisfied. In particular, there is often a region within which orbits are
trapped by the resonance. Trapped orbits are qualitatively different from
their nearby untrapped brethren.

B16 investigated trapping of the orbits of halo stars by the resonance
$\Omega_r\simeq\Omega_z$ in an axisymmetric potential.
Perturbation theory was used to construct orbital tori that allow accurate
reconstructions of trapped orbits. Here we have extended this work to the
construction of tori that provide accurate representations of orbits that are
strongly affected by resonances with the Galactic bar. 

In the case of the bar, perturbative effects on tori need to be considered
not only in the case of trapping, but also when an orbit circulates
outside the trapping region, because \TM's fitting
routine ignores all
non-axisymmetric components of the potential, and in contrast to the
situation encountered in B16, its tori do not provide adequate fits to all
circulating orbits.

We have used resonant perturbation theory to construct orbital tori for a
realistic representation of our barred Galaxy. The constructed tori provide
accurate representations of numerically integrated orbits because (a) they
employ the sophisticated pendulum equation introduced by
\cite{Kaasalainen_res} rather than the standard pendulum equation, and (b) we
employ the angle-action coordinates provided by \TM\ rather than the epicycle
approximation.

Having orbital tori rather than just numerically integrated orbits is
valuable for several reasons.  First, orbital tori provide an extremely
compact representation of orbits: several dozen numbers encode everything you
could possible wish to compute about the infinite number of orbits that can
be constructed by interpolating between tori.  Moreover, tori are quantified
by actions, which are adiabatically invariant constants of motion that also
allow one to compute the phase-space volumes occupied by sets of orbits.
Other substantial advantages of using tori are the ability to use an
analytic distribution function to populate orbits in a considered manner,
and  the ability to determine whether a given orbit will ever reach a
particular location, such as the immediate solar neighbourhood, and if so
with what velocities. Finally, tori make it possible to relax the potential
in which a multi-component system is constructed to self-consistency.

We have illustrated this power of orbital tori by using them to construct a
slice through velocity space at the Sun. To minimise the impact of Poisson
noise in this slice, action space had to be sampled in a non-uniform manner
because in some parts of velocity space the stellar density derives from a
small number of orbits that contribute with large weights.  This generic
drawback of orbit modelling can be side-stepped by building a map of
the value of each action velocity space from the value it takes
on an irregular distribution of points contributed by individual tori. 

\section*{Acknowledgements}

I thanks the referee Martin Weinberg for insightful comments.
This work was supported by
the European Research Council under the European Union's Seventh Framework
Programme (FP7/2007-2013)/ERC grant agreement no.~321067.

\bibliographystyle{mn2e} \bibliography{/u/tex/papers/mcmillan/torus/new_refs}

\appendix

\section{New classes for TM}\label{sec:TMclass}

To implement the algorithms presented here, four new classes, {\tt eTorus},
{\tt iTorus}, {\tt resTorus\_c} and {\tt resTorus\_L}, have been introduced
to the Torus Mapper \citep{JJBPJM16}.  Objects of type {\tt eTorus} are just
tools to be exploited by objects in the other three classes. Whereas an
object of class {\tt iTorus} uses non-resonant perturbation theory so provide
non-axisymmetric tori away from resonances, objects
of the last two classes use resonant perturbation theory and provide tori
near resonances.

\subsection{Class eTorus}

An {\tt eTorus} comprises an axisymmetric {\tt Torus} together with the
eight largest Fourier coefficiets $h_\vk$ of the full Hamiltonian on that
torus. The principal methods in this class are listed in
Table~\ref{tab:Etorus}.

\subsection{Class iTorus}

An {\tt iTorus} is an object whose methods encode non-resonant perturbation
theory. It is characterised by actions $J_i$ that quantify generalisations of
the standard actions of axisymmetric orbits, and it employs a grid of {\tt
eTori} to compute its properties.  {\tt FullMap} returns the phase-space
location given angle values. {\tt containsPoint} returns the number of
distinct velocities (if any) at which the torus visits a given spatial
location, along with the corresponding values of the angles and the Jacobian
$\p(\vx)/\p(\vtheta)$. The Boolean {\tt InOrbit} simply determines whether a
given point is visited. For given $J_r$ and $J_z$ {\tt get\_crit\_Jp} returns
the values of $J_\phi$ at which an orbit has apo- or peri-centre at a given
location. {\tt SOS} produces an $(R,p_R)$ surface of section.

Here is the code that created the $20\times30$ grid of eTori used to
produce figures that required untrapped orbits. Notice that the grid is
uniform in $\surd J_r$ rather than in $J_r$. {\tt Phi} and and {\tt bar} are
pointers to an
axisymmetric potential and an $m=2$ perturbation, respectively. 

{\obeylines\tt\parindent=10pt
	int nr=30,np=20;
	Actions J,Jgrid,dJ;
	dJ[0]=(sqrt(Jrmax)-sqrt(Jrmin))/(double)(nr-1); 
	dJ[1]=0; dJ[2]=(Jpmax-Jpmin)/(double)(np-1);
	Jgrid[0]=sqrt(Jrmin)+.5*dJ[0];  
	Jgrid[1]=J[1]; 
	Jgrid[2]=Jpmin+.5*dJ[2];
	eTorus **Tgrid=PJM::matrix<eTorus>(np,nr);
	strcat(fname,"nores.ebf"); bool writeit=true;
	if(writeit)\{
\qquad		ebf::Write(fname,"/dJ",\&dJ[0],"w","",1);
\qquad		printf("Computing eTorus grid:$\backslash$n");
\qquad		for(int i=0;i<np;i++)\{
\qquad\qquad		J[2]=Jpmin+i*dJ[2];
\qquad\qquad		for(int j=0;j<nr;j++)\{
\qquad\qquad\qquad		J[0]=pow(sqrt(Jrmin)+j*dJ[0],2);
\qquad\qquad\qquad		Tgrid[i][j].AutoFit(J,Phi,bar,Omp,tolJ);
\qquad\qquad\qquad		char lab[7]; 
\qquad\qquad\qquad		sprintf(lab,"eT\%d-\%d",i,j); 
\qquad\qquad\qquad		const string tname(lab);
\qquad\qquad\qquad		Tgrid[i][j].write\_ebf(fname,tname);
\qquad\qquad		\}
\qquad		\}
	\}else\{
\qquad		ebf::Read(fname,"/dJ",\&dJ[0],1);
\qquad		printf("Reading eTorus grid:$\backslash$n");
\qquad		for(int i=0;i<np;i++)\{
\qquad\qquad		for(int j=0;j<nr;j++)\{
\qquad\qquad\qquad		char lab[7];sprintf(lab,"eT\%d-\%d",i,j); 
\qquad\qquad\qquad		string tname(lab);
\qquad\qquad\qquad		Tgrid[i][j].read\_ebf(fname,tname);
\qquad\qquad		\}
\qquad		\}
	\}
}

\noindent Now 

{\obeylines\tt\parindent=10pt
		iTorus T(J,Tgrid,np,nr,Jgrid,dJ);
}

\noindent will create an untrapped torus with actions {\tt J}, which can be examined
with {\tt T.FullMap}, {\tt T.containsPoint} , etc

\begin{table*}
\caption{Public methods of an {\tt eTorus}}\label{tab:Etorus}
\begin{tabular}{ll}
\hline
{\tt eTorus()}&Null constructor\\
{\tt
eTorus(Actions,Potential*,bar\_pot*,double,double)}&{\hsize=.6\hsize\vtop{\noindent
Constructs an {\tt
eTorus} for given actions and potential. The last two arguments are the pattern speed and the tolerance
parameter {\tt tolJ}}}\\
{\tt eTorus(Torus\&,Potential*,bar\_pot*,double)}&Upgrades a {\tt Torus} to an
{\tt eTorus}. the last argument is the pattern speed\\
{\tt AutoFit(Actions,Potential*,bar\_pot*,double,double}&Changes an already
existing {\tt eTorus} so it has given actions.\\
{\tt reset(Torus\&,Potential*,bar\_pot*)}&Changes the home {\tt Torus} and on
it computes the $h_\vk$\\
{\tt Frequencies()}& Returns frequencies\\
{\tt actions()}&Returns actions\\
{\tt hn()}&Returns the values of the Fourier coefficients $h_\vk$\\
{\tt i1(), j1(), k1()}&Return  the integer coefficient of $\theta_r$,
$\theta_z$ and $\theta_\phi$ associated with
each term in $h_\vk$\\ 
{\tt FullMap(Angles)}&Returns $(R,z,\phi,v_R,v_z,v_\phi)$ of phase-space point
referenced by the given angles\\
{\hsize=.35\hsize\vtop{{\tt
containsPoint(Position\&,Velocity\&,Velocity\&,
double\&,Angles,Angles\&\&Velocity\&,
Velocity\&,double\&,Angles\&,Angles\&)}}}&
{\hsize=.56\hsize\vtop{Places in the  Angles arguments the angles at which the torus reaches the
given Position. The Velocity and double arguments are returned with the
velocities and inverse densities of the visits.}}\\
{\tt write\_ebf(string,string)}&{\hsize=.56\hsize\vtop{Writes the {\tt eTorus} to the ebf file named
by the first string with tag given by the second string}}\\
{\tt read\_ebf(string,string)}&Inverse of the above write method\\
\hline
\end{tabular}
\end{table*}

\begin{table*}
\caption{Public methods of an {\tt iTorus}}\label{tab:Itorus}
\begin{tabular}{ll}
\hline
{\tt iTorus(Actions,eTorus**,int,int,Actions,Actions}&Creates an {\tt iTorus}
with the given actions
given a grid of {\tt eTorus}s\\
{\tt eT1(Actions)}&Returns the {\tt eTorus} with the given actions\\
{\tt hn(), i1(), j1(), k1()}&Return properties of the home {\tt eTorus}\\
{\tt actions(), omega()}&Return actions and frequencies of home {\tt
eTorus}\\
{\tt FullMap(Angles)}&Returns $(R,z,\phi,v_R,v_z,v_\phi)$ pointed to by given angle\\
{\tt SOS(ostream\&,int)}&Sends surface of section to output stream\\
{\hsize=.35\hsize\vtop{{\tt
containsPoint(Position\&,Velocity*,Angles*, double*,int)}}}&{\hsize=.56\hsize\vtop{Returns number of
angles (up to maximum specified by last argument) at which Position is visited and leaves angles, velocities and inverse
densities of visits in arguments}}\\
\hline
\end{tabular}
\end{table*}

\subsection{Class resTorus\_c}

Objects in this class use resonant perturbation theory to compute tori that
are trapped, or nearly trapped, at corotation. The principal methods are
listed in Table~\ref{tab:resTorusc}. The constructor requires values of
$J_r,\,J_z$ and a rough estimate of the value of $J_\phi$ for corotation
resonance with the given $(J_r,J_z)$ -- the estimate of $J_\phi$ is changed
to achieve precise corotation resonance.  An {\tt eTorus} with these actions
is constructed and the parameters of the pendulum equation found by Taylor
expanding the Hamiltonian's Fourier amplitudes around the values taken on the
perfectly resonant {\tt eTorus}. From these parameters the extent of the region of
entrapment is determined and then a grid of tori is computed so any trapped
or nearly trapped torus can be constructed by interpolation. 

A particular
trapped torus is specified by giving a value of the variable $I$ with {\tt
setI}. Given an instance {\tt resTorus\_c T}, the maximum permitted
value, {\tt
T.Imax}, generates the torus with no libration, while the value {\tt T.Imin}
generates the torus with maximum libration. Values smaller than {\tt T.Imin}
generate circulation. By default circulation takes place outside
corotation, but {\tt T.setI(1.05*T.Imin,-1)} will generate circulation inside
corotation.

\begin{table*}
\caption{Public methods of an {\tt resTorus\_c}}\label{tab:resTorusc}
\begin{tabular}{ll}
\hline
{\hsize=.35\hsize\vtop{{\tt
resTorus\_c(Torus**,int,Actions,Potential*, bar\_pot*,double,double)}}}&
{\hsize=.56\hsize\vtop{Constructor given actions and a pointer to an empty $2\times n$ grid of
tori. The second argument specifies $n$.}}\\
{\tt setI(double)}&Sets variable $I$ that controls the amplitude of
libration\\
{\tt setI(double,int)}&As {\tt setI(double)} but if $I<I_{\rm bot}$,
circulates inside corotation if integer is $-1$\\
{\tt librationAction()}&Returns action of libration.\\
{\tt librationOmega()}&Returns libration frequency\\
{\tt from\_librationAngle(double,double\&,double\&)}&Return $\theta_\phi$,
$\d\theta_\phi/\d\theta_{\rm la}$ and offset of $J_r$ from perfectly resonant
value\\
{\tt FullMap(Angles)}&Returns $(R,z,\phi,v_R,v_z,v_\phi)$ for given angles\\
{\tt SOS(ostream\&,int)}&Send $\phi=\pi/2$ surface of section to ostream\\
{\hsize=.35\hsize\vtop{{\tt
containsPoint(Position\&,Velocity*,Angles*, Angles*,double*)}}}&{\hsize=.6\hsize\vtop{Returns number of
angles at which Position is visited and leaves angles, velocities and inverse
densities of visits in arguments}}\\
{\tt getJs(Actions\&,Actions\&,double)}&Returns extremes of unperturbed
actions on orbit\\
{\tt get\_resJp()}&Returns actions of underlying perfectly resonant torus\\
{\tt omega()}&Returns frequencies\\
\hline
\end{tabular}
\end{table*}

With {\tt Phi} a pointer to an axisymmetric potential and {\tt bar} a pointer
to an $m=2$ perturbation, the following code will generate data for
Fig.~\ref{fig:circulate}

{\obeylines\tt\parindent=10pt
	FILE *ofile; ofile=fopen("corot.out","w");
	Actions J; J[0]=.1; J[1]=0.0025; J[2]=2;
	int np=5; double Omp=0.04,tolJ=0.003;
	Torus **Tgrid; Tgrid = PJM::matrix<Torus>(2,np);
	resTorus\_c T(Tgrid,np,J,Phi,bar,Omp,tolJ);
	T.setI(1.1*T.Imin,1); J=T.get\_resJp();
	Angles A; Frequencies Omres=T.omega();
	int NS=5000; double x[NS],y[NS];
	double Jl=T.librationAction();
	double dt=10*PI/fabs(Omres[2])/(double)NS;
	fprintf(ofile,"\%f \%f \%f$\backslash$n",(NS-1)*dt,J[0],Jl);
	for(int i=0;i<NS;i++)\{
\qquad		double t=i*dt;
\qquad		A[0]=Omres[0]*t; A[1]=Omres[1]*t;
\qquad		 A[2]=Omres[2]*t;
\qquad		GCY gcy=T.FullMap(A);
\qquad		x[i]=gcy[0]*cos(gcy[2]); 
\qquad		y[i]=gcy[0]*sin(gcy[2]);
\qquad		for(int j=0;j<6;j++)\{
\qquad\qquad			fprintf(ofile,"\%f ",gcy[j]);
\qquad		\}
\qquad		fprintf(ofile,"$\backslash$n");
	\}
}
Connecting the points $(x,y)$ stored in {\tt x[],y[]} will now produce the red
curve in Fig.~\ref{fig:circulate}, while an orbit integration can be launched
from any of the phase-space points written to `corot.out'.

\subsection{Class resTorus\_L}

Instances of this class generate tori that are trapped or nearly trapped at a
Lindblad resonance. The layout is very similar to that of {\tt resTorus\_c}
(Table~\ref{tab:resTorusL}).
One more argument is required, namely the integer 3-vector {\tt resN}, which
is $(1,0,2)$ for OLR and $(1,0,-2)$ for ILR. The following statements create
a torus with the values of $J_r$ and $J_z$ given in {\tt J} that is trapped at OLR

{\obeylines\tt\parindent=10pt 
int nr=5; Tgrid = PJM::matrix<Torus>(nr,2);
int sc[3]={1,0,2}; int3 resN(sc);
resTorus\_L T(Tgrid,nr,J,Phi,bar,Omp,resN,tolJ); 
T.setI(.8*T.Imin());
}

\noindent The properties of this torus can then be explored with {\tt T.FullMap},
{\tt T.containsPoint}, etc, exactly as for any other torus.

\subsection{Calling programs}

To illustrate the use of these tools the \TM\ depository contains three
main programs: {\tt corot.cc}, {\tt lindblad.cc} and {\tt nores.cc}. The
first two programs will:

\begin{enumerate}
\item Compute the edges of the ladders plotted in Figs.~\ref{fig:Hbar}
and \ref{fig:DF};
\item Compute the velocities at which a resonantly trapped torus visits a
Sun-like location;
\item Plot a trapped or near-trapped orbit and produce the data required for
a surface of section.
\end{enumerate}

The third main program reads files created by the first two and computes the
velocities at which non-resonant orbits reach a Sun-like location. It will
also produce data to plot such an orbit.

\begin{table*}
\caption{Public methods of an {\tt resTorus\_L}}\label{tab:resTorusL}
\begin{tabular}{ll}
\hline
{\hsize=.35\hsize\vtop{{\tt
resTorus\_L(Torus**,int,Actions,Potential*, bar\_pot*,double,int3,double)}}}&
{\hsize=.56\hsize\vtop{Constructor given actions and a pointer to an empty $2\times n$ grid of
tori. The second argument specifies $n$.}}\\
{\tt setI(double)}&Sets variable $I$ that controls the amplitude of
libration\\
{\tt setI(double,int)}&As {\tt setI(double)} but if $I<I_{\rm bot}$,
circulates inside corotation if integer is $-1$\\
{\tt librationAction()}&Returns action of libration.\\
{\tt librationOmega()}&Returns libration frequency\\
{\tt from\_librationAngle(double,double\&,double\&)}&Return $\theta_1'$,
$\d\theta_1'/\d\theta_{\rm la}$ and offset in $J_1'$ from the perfectly
resonant value\\
{\tt FullMap(Angles)}&Returns $(R,z,\phi,v_R,v_z,v_\phi)$ for given angles\\
{\tt SOS(ostream\&,int)}&Send $\phi=\pi/2$ surface of section to ostream\\
{\hsize=.35\hsize\vtop{{\tt
containsPoint(Position\&,Velocity*,Angles*, Angles*,double*)}}}&{\hsize=.6\hsize\vtop{Returns number of
angles at which Position is visited and leaves angles, velocities and inverse
densities of visits in arguments}}\\
{\tt getJs(Actions\&,Actions\&,double)}&Returns extremes of unperturbed
actions on orbit\\
{\tt get\_resJp()}&Returns actions of perfectly resonant {\tt eTorus}\\
{\tt omega()}&Returns frequencies\\
{\tt prime\_it(Action,bool)}&If the {\tt bool}=1, $J'\to J$, else reverse\\
\hline
\end{tabular}
\end{table*}

\section{Finding velocities at a given location}\label{sec:getV}

Whereas it is trivial to reach a given azimuth $\phi_0$ on an axisymmetric
torus (by adding whatever we like to $\theta_\phi$), on a non-axisymmetric
torus $\theta_\phi$ becomes a non-trivial variable in the sense that we
cannot change it without changing the toy angles $\theta^{\rm T}$, so the quantity to be
minimised by {\tt containsPoint} must have three terms
\[
\chi^2=(R-R_0)^2+(z-z_0)^2+R_0^2(\phi-\phi_0)^2
\]
 rather than just the first two. Hence we now need the derivatives of $\phi$
 with respect to $\vtheta^{\rm T}$ and $\vJ^{\rm T}$. Since
\[
\phi=u+\thetaT_\phi-\sgn(\JT_\phi)\thetaT_\vartheta
\]
with
\[
\sin u\equiv\cot i\tan\vartheta\hbox{ and }\cos i\equiv{\JT_\phi\over
\JT_\vartheta+|\JT_\phi|}\equiv {\JT_\phi\over L},
\]
we obtain
\begin{align}
{\p\phi\over\p\thetaT_r}&=\sec u{\cot i\over\cos^2\vartheta}{\p\vartheta\over\p\thetaT_r}\cr
{\p\phi\over\p\thetaT_\vartheta}&=\sec u{\cot
i\over\cos^2\vartheta}{\p\vartheta\over\p\thetaT_\vartheta}-\sgn(\JT_\phi)\cr
{\p\phi\over\p\thetaT_\phi}&=1.
\end{align}
 Here the last equation follows because $\p\vartheta/\p\thetaT_\phi=0$ and
$\vartheta$ is latitude rather than the conventional polar angle because \TM\
adopts this notation. Similarly,
\begin{align}
 {\p\phi\over\p \JT_r}&={\p u\over\p \JT_r}
=\sec u{\cot i\over\cos^2\vartheta}{\p\vartheta\over\p \JT_r}\cr
{\p\phi\over\p \JT_\vartheta}&={\p u\over\p \JT_\vartheta}
=\sec u\left({\cot i\over\cos^2\vartheta}{\p\vartheta\over\p \JT_\theta}-{\cot
\vartheta\over\sin^3 i}{\JT_\phi\over L^2}\right)\cr
{\p\phi\over\p \JT_\phi}&={\p u\over\p \JT_\phi}
=\sec u\left({\cot i\over\cos^2\vartheta}{\p\vartheta\over\p
\JT_\phi}+{\cot\vartheta\over\sin^3 i}{\JT_\vartheta\over L^2}\right)
\end{align}

\TM\ finds the velocities at which an axisymetric torus visits a given
location $\vx$ by varying the toy angle variables, which
are most directly related to ordinary phase-space coordinates. When the torus
is not axisymmetric this is not a good strategy, not least because on a
trapped torus many values of $\thetaT$ are inaccessible. So {\tt iTorus},
{\tt resTorus\_c} and {\tt resTorus\_L} all search in true angles.

Let $(\vtheta',\vJ')$ denote the angle-action coordinates of a non-axisymmetric
torus,  $(\vtheta,\vJ)$ denote angle-action coordinates for a realistic
axisymmetric Galactic potential, and $(\vtheta^{\rm T},\vJ^{\rm T})$ denote the `toy'
angle-action cordinates of an isochrone potential. Then $(\vtheta,\vJ)$ and
$(\vtheta^{\rm T},\vJ^{\rm T})$ are related by \TM's standard generating function
\[
S(\vtheta^{\rm T},\vJ)=\vtheta^{\rm T}\cdot\vJ+2\sum_\vn S_\vn(\vJ)\sin(\vn\cdot\vtheta^{\rm T}).
\]
$S$ and its derivatives are provided by \TM. 
We use the Newton-Raphson algorithm to find the values of $\vtheta'$ at which
the non-axisymmetric torus $\vJ'$ passes through the point $\vx$. We have
\[
\biggl({\p\vx\over\p\vtheta'}\biggr)_{\vJ'}=
\biggl({\p\vx\over\p\vthetaT}\biggr)_{\vJT}\biggl({\p\vthetaT\over\p\vtheta'}\biggr)_{\vJ'}
+
\biggl({\p\vx\over\p\vJT}\biggr)_{\vthetaT}\biggl({\p\vJT\over\p\vtheta'}\biggr)_{\vJ'}
\]
and
\[
\biggl({\p\vJT\over\p\vtheta'}\biggr)_{\vJ'}=
\biggl({\p\vJT\over\p\vJ}\biggr)_{\vthetaT}\biggl({\p\vJ\over\p\vtheta'}\biggr)_{\vJ'}
+
\biggl({\p\vJT\over\p\vthetaT}\biggr)_{\vJ}\biggl({\p\vthetaT\over\p\vtheta'}\biggr)_{\vJ'}
\]
so
\begin{align}
\biggl({\p\vx\over\p\vtheta'}\biggr)_{\vJ'}=&
\biggl\{\biggl({\p\vx\over\p\vthetaT}\biggr)_{\vJT}+
\biggl({\p\vx\over\p\vJT}\biggr)_{\vthetaT}\biggl({\p\vJT\over\p\vthetaT}\biggr)_{\vJ}
\biggr\}
\biggl({\p\vthetaT\over\p\vtheta'}\biggr)_{\vJ'}\cr
&+\biggl({\p\vx\over\p\vJT}\biggr)_{\vthetaT}
\biggl({\p\vJT\over\p\vJ}\biggr)_{\vthetaT}\biggl({\p\vJ\over\p\vtheta'}\biggr)_{\vJ'}
\end{align}
 The last three derivatives are available from (i) the toy action-angle
relations, (ii) the generating function, and (iii) perturbation theory. The
derivatives in the curly bracket of the line above are available from
the toy angle-action coordinates and the generating function. The 
derivative that multiplies the curly bracket is problematic, however.
Specifically
\[\label{eq:B4}
\biggl({\p\vthetaT\over\p\vtheta'}\biggr)_{\vJ'}=
\biggl({\p\vthetaT\over\p\vJ}\biggr)_{\vtheta}\biggl({\p\vJ\over\p\vtheta'}\biggr)_{\vJ'}
+
\biggl({\p\vthetaT\over\p\vtheta}\biggr)_{\vJ}\biggl({\p\vtheta\over\p\vtheta'}\biggr)_{\vJ'}
\]
The second and last derivatives on the right here are provided by
perturbation theory, while the first and third derivatives are in principle
available from the generating function. However, the first derivative must be
obtained by differentiating $\vtheta=\p S/\p\vJ$, which yields
\[
0={\p^2S\over\p J_k\p J_j}
+{\p^2S\over\p\thetaT_i\p J_j}\biggl({\p\thetaT_i\over\p
J_k}\biggr)_{\vtheta}.
\]
 Unfortunately, we have no way of computing the first of these double
derivatives of $S$. Fortunately, the first term on the right of equation
(\ref{eq:B4}) is of order the perturbation, whereas  the second is not. So
we neglect the first term.

\end{document}